\documentclass[iop,twocolappendix]{emulateapj}
\usepackage{blindtext}
\usepackage{placeins}
\usepackage{graphicx}
\usepackage{amssymb, amsmath}
\usepackage{natbib}
\usepackage{amsmath}
\usepackage{bm}
\usepackage{color}
\usepackage[colorlinks,urlcolor=blue,citecolor=blue,linkcolor=blue]{hyperref}
\usepackage[caption=false]{subfig}

\begin{document}
\title{The Mechanism of Electron Injection and Acceleration in trans-relativistic reconnection}

\author{David Ball,\altaffilmark{1,3} Lorenzo Sironi,\altaffilmark{2} and Feryal \"Ozel\altaffilmark{1}}

\altaffiltext{1}{Department of Astronomy and Steward Observatory, Univ. of Arizona, 933 N. Cherry Avenue, Tucson, AZ 85721, USA}

\altaffiltext{2}{Department of Astronomy, Columbia University, 550 West 120th Street, New York, NY 10027, USA}
\altaffiltext{3}{Email: davidrball@email.arizona.edu}

\begin{abstract}

Electron acceleration during magnetic reconnection is thought to play a key role in time-variable high-energy emission from astrophysical systems.  
By means of particle-in-cell simulations of trans-relativistic reconnection, we investigate electron injection and acceleration mechanisms in low-$\beta$ electron-proton plasmas.  We set up a diversity of  density and field structures (e.g., X-points and plasmoids) by varying the guide field strength and choosing whether to trigger reconnection or let it spontaneously evolve.  We show that the number of X-points and plasmoids controls the efficiency of electron acceleration, with more X-points leading to a higher efficiency.  Using on-the-fly acceleration diagnostics, we also show that the non-ideal electric fields associated with X-points play a critical role in the first stages of electron acceleration.  As a further diagnostic, we include two populations of test particles that selectively experience only certain components of electric fields.  We find that the out-of-plane component of the parallel electric field determines the hardness of the high-energy tail of the electron energy distribution.  These results further our understanding of electron acceleration in this regime of magnetic reconnection and have implications for realistic models of black hole accretion flows.

\end{abstract}
\maketitle

\section{introduction} \label{introduction}
Magnetic reconnection is thought to play an important role in accelerating electrons and powering high-energy emission in numerous astrophysical systems including accretion flows around black holes (\citealt{galeev1979}; \citealt{dimatteo1998}; \citealt{uzdensky2008}; \citealt{li2015}; \citealt{ball2016, ball2017}; \citealt{li2017}; \citealt{dalpino2018}; \citealt{ramirez2018}), blazar jets (\citealt{romanova1992}; \citealt{giannios2009}; \citealt{giannios2010}; \citealt{giannios2013}; \citealt{petropoulou2016}; \citealt{nalewajko2016}; \citealt{nalewajko2018b}; \citealt{christie2019}), pulsars and pulsar wind nebulae (\citealt{coroniti1990}; \citealt{lyubarsky2001}; \citealt{zenitani2001}; \citealt{kirk2003}; \citealt{contopoulos2007,contopoulos2007b}; \citealt{zenitani2007}; \citealt{petri2008}; \citealt{sironi2011}; \citealt{cerutti2012, cerutti2014}; \citealt{cerutti2017}; \citealt{philippov2014}; \citealt{hakobyan2018}), gamma ray bursts (\citealt{thompson1994, thompson2006}; \citealt{usov1994}; \citealt{spruit2001}; \citealt{drenkhahn2002}; \citealt{lyutikov2003}; \citealt{giannios2008}; \citealt{beniamini2017}; \citealt{werner2019}), and the Sun (\citealt{forbes1996}; \citealt{yokoyama2001}; \citealt{shibata2011}).  Despite its ubiquity, the physics of particle acceleration in reconnection is not fully understood.  

The nature of reconnection depends on a few key properties of the plasma.  We refer to the ratio of magnetic energy density to plasma enthalpy density as the ``magnetization'',
$\sigma=B_{0}^2 / 4\pi w_{0} $, where $B_{0}$ is the magnetic field strength and $w_0=(\rho_{e}+\rho_{i})c^{2}+ \hat{\gamma}_{e}u_{e}+ \hat{\gamma}_{i}u_{i}$ is the plasma enthalpy density.  Here, $\rho_{i,e}$, $\hat{\gamma}_{i,e}$, and $u_{i,e}$ are the mass densities, adiabatic indices, and internal energy densities of ambient protons and electrons, respectively.  This parameter
controls the bulk energization of the plasma and the efficiency of particle acceleration.  When $\sigma$ is of order unity, we refer to the plasma as ``trans-relativistic''.  In this regime, post-reconnection protons remain sub-relativistic ($\gamma_{i} \approx 1$), while electrons can be heated and accelerated to ultra-relativistic energies ($\gamma_{e} \gg 1$). 
The plasma-$\beta$, i.e., the ratio of gas pressure to magnetic pressure also plays an important role in the dynamics of reconnection and controls --- together with $\sigma$ --- the shape of the electron energy spectrum (\citealt{ball2018}).  In this work, when $\beta$ is small, we refer to the plasma as ``magnetically dominated''.  Trans-relativistic magnetically dominated plasmas occur frequently in radiatively inefficient accretion flows such as Sgr~A* and the supermassive black hole at the center of M87, particularly in the coronae and strongly magnetized regions close the black hole's event horizon.   

Numerous PIC studies have investigated electron acceleration mechanisms in both relativistic (\citealt{zenitani2001, zenitani2007}; \citealt{lyubarsky2008}; \citealt{sironi2014}; \citealt{melzani2014b,melzani2014}; \citealt{nalewajko2015}; \citealt{guo2015, guo2019}; \citealt{werner2016, werner2017}; \citealt{petropoulou2018}) and nonrelativistic reconnection (\citealt{dahlin2014, dahlin2016, dahlin2016b, wangh2016, li_guo_2017}).  

Some of these studies apply a guiding center formalism (e.g., \citealt{dahlin2014}).  In doing so, they can separate different terms in the energy equation, and by summing over particles, they can ultimately assess the various contributions to bulk energization. Such a treatment will properly capture the physics of bulk heating, but will not highlight the relatively small number of accelerated particles in a high-energy non-thermal tail.  Additionally, the guiding center formalism breaks down at X-points in anti-parallel reconnection, where the magnetic field vanishes.  Others look at individual particle trajectories to assess where and by what mechanisms particles are being accelerated.  In most cases these studies sparsely sample a collection of representative particles.  Furthermore, most of these studies employ either a pair plasma or a significantly reduced mass ratio, which may affect the conclusions. 

These previous studies have highlighted a few distinct acceleration mechanisms.  One is acceleration by the non-ideal out-of-plane electric field (i.e., in the direction of the electric current) at X-points (see e.g., \citealt{zenitani2001, sironi2014, nalewajko2015}).  These X-points can occur not only in the initial current sheet via the primary or secondary tearing mode, but also in the current sheets generated between merging plasmoids.  Another prominent mechanism is Fermi reflection (see, e.g., \citealt{dahlin2016b}; \citealt{guo2019}), enabled by the various macro-scale motions induced by reconnection which can occur within contracting plasmoids (\citealt{drake2006}) and also between outflows and plasmoids (\citealt{ball2018}).  \citet{nalewajko2015} also found that particles can be accelerated in the trailing edges of accelerating plasmoids.  \citet{petropoulou2018} found that the particles dominating the high-energy spectral cut-off reside in plasmoids, and their acceleration is driven by the increase in the local field strength as plasmoids grow and compress, coupled with the conservation of the first adiabatic invariant.  More recently, \citet{guo2019} found that for a relativistic ($\sigma=50$) pair plasma the ideal electric field and associated Fermi reflection is sufficient to produce a non-thermal distribution extending to high energies, i.e., they argued for a negligible role of non-ideal electric fields.

In \citet{ball2018} we investigated particle acceleration in the trans-relativistic regime, where the plasma magnetization (i.e., ratio of magnetic energy density to particle enthalpy density) is of order unity,  and found preliminary evidence for the role of X-points, plasmoids, and overall bulk motions in promoting electron acceleration by examining the histories of a few representative high-energy electrons.  We did not, however, systematically examine an unbiased sample of electrons.  In that study, we observed that at low-$\beta$, electrons undergo extremely short periods of intense acceleration by a non-ideal electric field at X-points in the initial sheet or between merging plasmoids.  In contrast, at high-$\beta$ ($\beta \sim 1/4\sigma$) when electrons start out relativistically hot, X-point acceleration is negligible, and Fermi reflection dominates.  The reason for this is twofold: First, the secondary tearing mode is suppressed in thermally dominated plasmas and so the chance of interacting with an X-point is extremely low; second, the energy gain via Fermi reflection off of structures moving at the Alfv\'en speed dominates over the energy gain at X-points when the electrons already start with relativistic velocities.

In this paper, we systematically investigate the physics of electron acceleration in trans-relativistic reconnection by comparing two-dimensional simulations in which we vary the number of X-points and plasmoids by changing (\textit{i}) the guide field strength and (\textit{ii}) whether or not we induce reconnection by hand or let it evolve naturally via the primary tearing instability.  We use the true electron-proton mass ratio in all of our simulations to ensure that our results are not affected by the choice of a non-physical reduced mass ratio.  We include particle acceleration diagnostics that are calculated on-the-fly during the simulation for all of the particles.  We specify the strength of the guide field, $B_{g}$, as a ratio of the reconnecting component, $B_{0}$, and choose values of $B_{g}/B_{0}=0.1$ and $B_{g}/B_{0}=0.3$.  We choose $B_{g}/B_{0}=0.1$ because it is remarkably similar to the purely anti-parallel ($B_{g}=0$) case in terms of the fluid structures and electron energy spectra, yet allows us to define the parallel electric field ($E_{||}=\bold{E} \cdot \hat{b}$) and corresponding work done by non-ideal parallel electric fields.  We choose $B_{g}/B_{0}=0.3$ as our other choice because at and above this threshold, the secondary tearing mode is suppressed, simplifying the physics and allowing us to more cleanly isolate the effects of X-point acceleration.

We find that X-points formed both in the primary current sheet (PCS) as well as in merger-induced current sheets (MCS) play critical roles in the first stages of electron acceleration in the trans-relativistic regime.  In particular, we find that when the secondary tearing mode is active, electron acceleration is enhanced because X-points occur more frequently throughout the current layer.  In these cases, X-points are ubiquitous, both in the PCS and MCS.  In contrast, when the secondary tearing mode is suppressed, high-energy acceleration is localized to the primary X-point(s), which are very few relative to the copious X-points that occur when secondary tearing is active.  We show that the first stages of electron acceleration are controlled by the out-of-plane component of the parallel electric field (i.e., by the non-ideal reconnection electric field).  After this initial acceleration, electrons are further energized via other processes controlled by ideal fields, that may eventually be dominant as compared to the X-point phase.  We further illustrate the importance of this pre-acceleration stage at X-points by tracking populations of test particles (i.e., which are moved by the Lorentz force in the same way as the ``real'' particles in our simulations but do not contribute to electric currents) that only feel certain components of the parallel electric field.  We show that the out-of-plane component of parallel electric fields is critical in producing a hard non-thermal tail that extends to high energies, as observed for the ``real'' electrons in our simulations.

The layout of the paper is as follows: in Section \ref{methods} we describe the simulations we employ, in Section \ref{diagnostics_of_acceleration} we explain the various diagnostics that we calculate on-the-fly for all of the particles in our simulation in order to understand their acceleration histories.  In Section \ref{role_of_xpoint} we show the results from our simulations and investigate the importance of X-points in accelerating high-energy electrons.  In Section \ref{E_comps} we investigate the role of the out-of-plane component of the parallel electric field by using our on-the-fly diagnostic of the cumulative work done on each particle by this component of the electric field.  Finally, in Section \ref{test_prtls}, we use test particles that do not feel certain components of the parallel electric field to show that parallel electric fields play crucial roles in regulating both the heating and acceleration of electrons. 
\\
\\

\section{methods} \label{methods}

\subsection{Simulation Setup} \label{sim_details}
We perform four fiducial simulations of magnetic reconnection using the publicly-available code TRISTAN-MP (\citealt{buneman93}; \citealt{spitkovsky05}).  As in \citet{ball2018}, we employ a two-dimensional (2D) simulation domain in the $xy$ plane, but we track all three components of velocity and electromagnetic field vectors.  The system is periodic in the $x$-direction and enlarged continuously in the $y$-direction as the simulation progresses (for details on the expanding box, see \citet{sironi2014}).  We set up the system in Harris equilibrium, with an in-plane magnetic field profile $\bm{{B}} = -B_{0}\tanh{\left(2\pi y / \Delta\right)}\bm{\hat{x}}$, where $B_{0}$ is the strength of the reconnecting field in the ambient plasma and $\Delta$ is the thickness of the sheet.  $B_{0}$ is related to the magnetization parameter $\sigma$ via $\sigma=B_{0}^2 / 4\pi w_{0} $.  We set $\sigma=0.3$ for all of our simulations (see \citet{werner2018}, \citet{ball2018} for a study of the dependence of electron acceleration on $\sigma$).  To achieve pressure equilibrium, we initialize the current sheet with a population of hot particles that are overdense compared to the upstream plasma by a factor of $\eta$, such that their temperature is given by $kT_{H}/m_{i}c^{2}=\sigma/2\eta$ (in the cold limit where the enthalpy density is approximately equal to the rest mass energy density).  We set $\eta=3$ for all the simulations in this paper.  Given that the properties of the hot particles initialized in the sheet depend on arbitrary choices at initialization, we exclude them from the particle spectra and from all our analyses.

\begin{figure*}[t]
	\includegraphics[width=\linewidth]{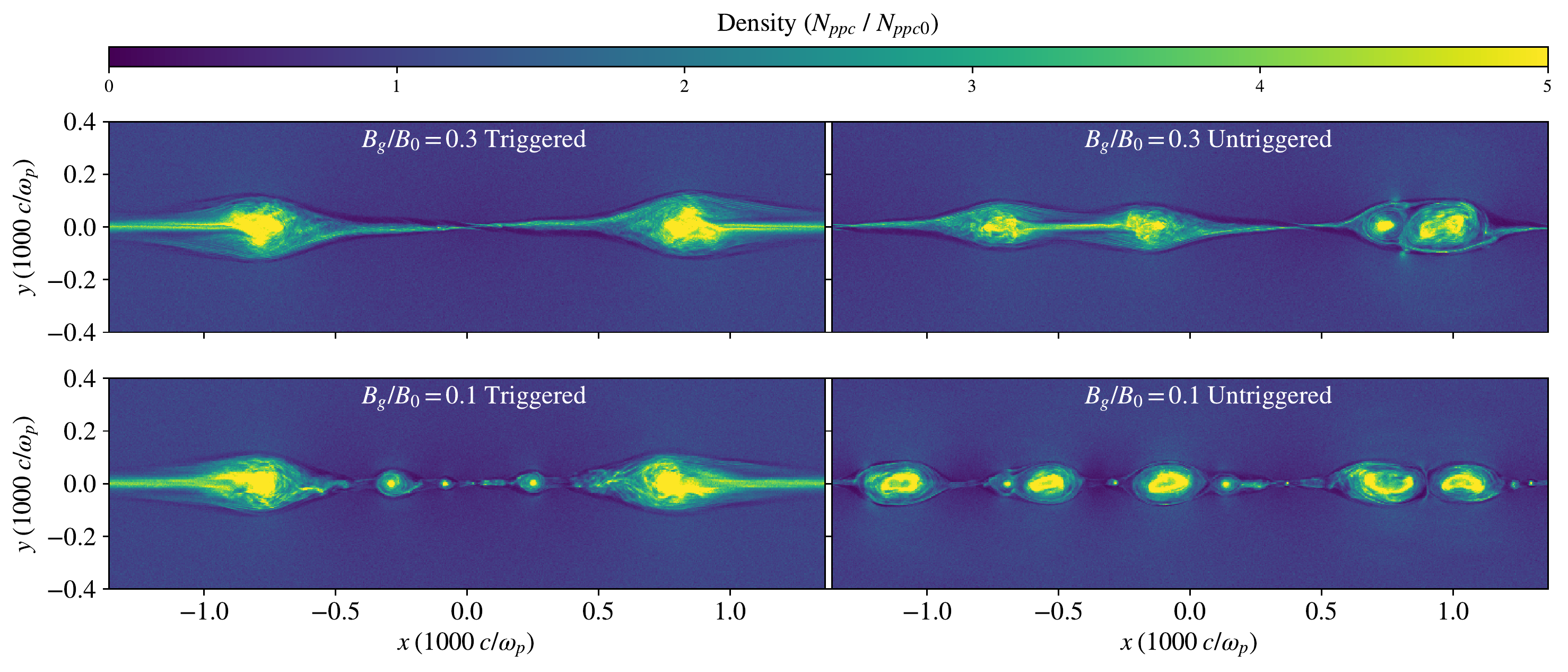}
	\caption{Snapshots of density from four fiducial simulations showing a diversity of configurations with different numbers of X-points and plasmoids.  All snapshots are taken at $t=3600 \; \omega_{p}^{-1}$, or in terms of Alfv\'en crossing times, $t_{A}=L/v_{A}$, $t \sim 0.5 t_{A}$.  The top row shows the simulations with a guide field strength of $B_{g}=0.3B_{0}$ and the bottom row shows the simulations with a guide field strength of $B_{g}=0.1B_{0}$.  The first column shows simulations where reconnection is triggered, and the second column shows simulations where reconnection develops spontaneously (i.e., untriggered).
	}
	\label{lowbeta_fourdens}
\end{figure*}

\begin{deluxetable*}{cccccc}
	\centering
	\tablewidth{1.5\columnwidth}
	\tablecaption{Simulation Parameters}  
	\tablehead{run&$B_{g}/B_{0}$& Triggered vs. Untriggered  & $L_{x}$ ($1000 \; c/\omega_{p}$) & $\Delta \; (c/\omega_{p})$ & Final time ($\omega_{p}^{-1}$)}
	\startdata
	A0*& 0.3 & Triggered& 2.7 &26.66& 19,500\\ 
	\hline
	A1& 0.3 & Triggered& 1.3 & 26.66& 10,000\\
	\hline
	A2& 0.3 & Triggered& 5.4 & 26.66& 39,000\\
	\hline
	
	B0* & 0.3&Untriggered & 2.7 & 13.33& 19,500\\ 
	\hline
	B1& 0.3 & Untriggered& 1.3 & 13.33& 10,000\\
	\hline
	B2& 0.3 & Untriggered& 5.4 & 13.33& 39,000\\
	\hline
	B3& 0.3 & Untriggered& 5.4 & 6.66& 39,000\\
	\hline
	B4& 0.3 & Untriggered& 5.4 & 20.0& 60,000\\
	\hline
	C0*&  0.1&Triggered& 2.7 & 26.66& 19,500\\ 
	\hline
	C1& 0.1 & Triggered& 1.3 & 26.66& 10,000\\
	\hline
	C2& 0.1 & Triggered& 5.4 & 26.66& 39,000\\
	\hline
	
	D0*&  0.1&Untriggered & 2.7 & 13.33& 19,500\\ 
	\hline
	D1& 0.1 & Untriggered& 1.3 & 13.33& 10,000\\
	\hline
	D2& 0.1 & Untriggered& 5.4 & 13.33& 39,000\\
	\hline
	E0& 0 & Triggered& 2.7 & 26.66& 19,500\\
	\enddata
	\tablecomments{Summary of the physical and numerical parameters of our simulations.  All simulations are performed with the physical electron-proton mass ratio, equal electron and proton temperatures, a resolution of 3 cells per electron skin depth, $\sigma=0.3$ and $\beta_{i}=0.003$.  Our four fiducial simulations are A0*, B0*, C0*, and D0*.  Simulations with the same guide field strength and triggering choice (but varying box size, current sheet width, etc.) are represented with the same letter but a different number.}
	\label{tab:fit}
\end{deluxetable*}

We include an out-of-plane magnetic field (referred to as a guide field) and specify its strength as a fraction of the initial in-plane component, $B_{g}/B_{0}$.  We use two values of guide field strength, $B_{g}/B_{0}=0.1$ and $B_{g}/B_{0}=0.3$.  We specify the temperature through the proton plasma beta, defined as $\beta_{i}= 8 \pi n_{i} k T_{i}/B_{0}^{2}$, where $n_i=\rho_i/m_i$ is the proton number density, $T_i$ is the proton temperature, and $m_i$ is the proton mass. Ambient electrons and protons start with the same temperature, so that $\beta_e=\beta_i$ (and the total plasma-beta, including both species, is $2\beta_{i}$).  We focus for this work on low beta reconnection, choosing a representative case with $\beta_{i}=0.003$, which in \citet{ball2018} we demonstrated to lead to efficient electron acceleration.  This $\beta_{i}$ corresponds to $\theta_{i}=5\times 10^{-4}$ and $\theta_{e}=0.918$, where $\theta_{s}$ is the dimensionless temperature of a species given by $\theta_{s}=kT_{s}/m_{s}c^{2}$: at this $\beta_{i}$ and $\sigma$, the ambient protons are non-relativistic.  Because of this, the magnetization parameter as defined with the proton rest mass $\sigma_{i}=B_0^2/4\pi \rho_i c^2$ (where $\rho_i$ is the mass density of protons) is nearly identical to the enthalpy-weighted magnetization $\sigma$ defined above. 

For all of our simulations, we use the true proton-electron mass ratio, $m_{i}/m_{e}=1836$.  We initialize each computational cell in the ambient plasma with four particles per cell ($N_{ppc0}=4$).  We resolve the electron skin depth of the ambient plasma with three computational cells, and the length of our domain, along the current sheet, is 8,160 cells, corresponding to 2,720 electron skin depths (see \citet{ball2018} for tests of numerical convergence).  Additionally, we have simulations at twice the box length for each set of physical parameters, which we explore in Appendix \ref{box_size}.

\begin{figure*}[t]
	\includegraphics[width=\linewidth]{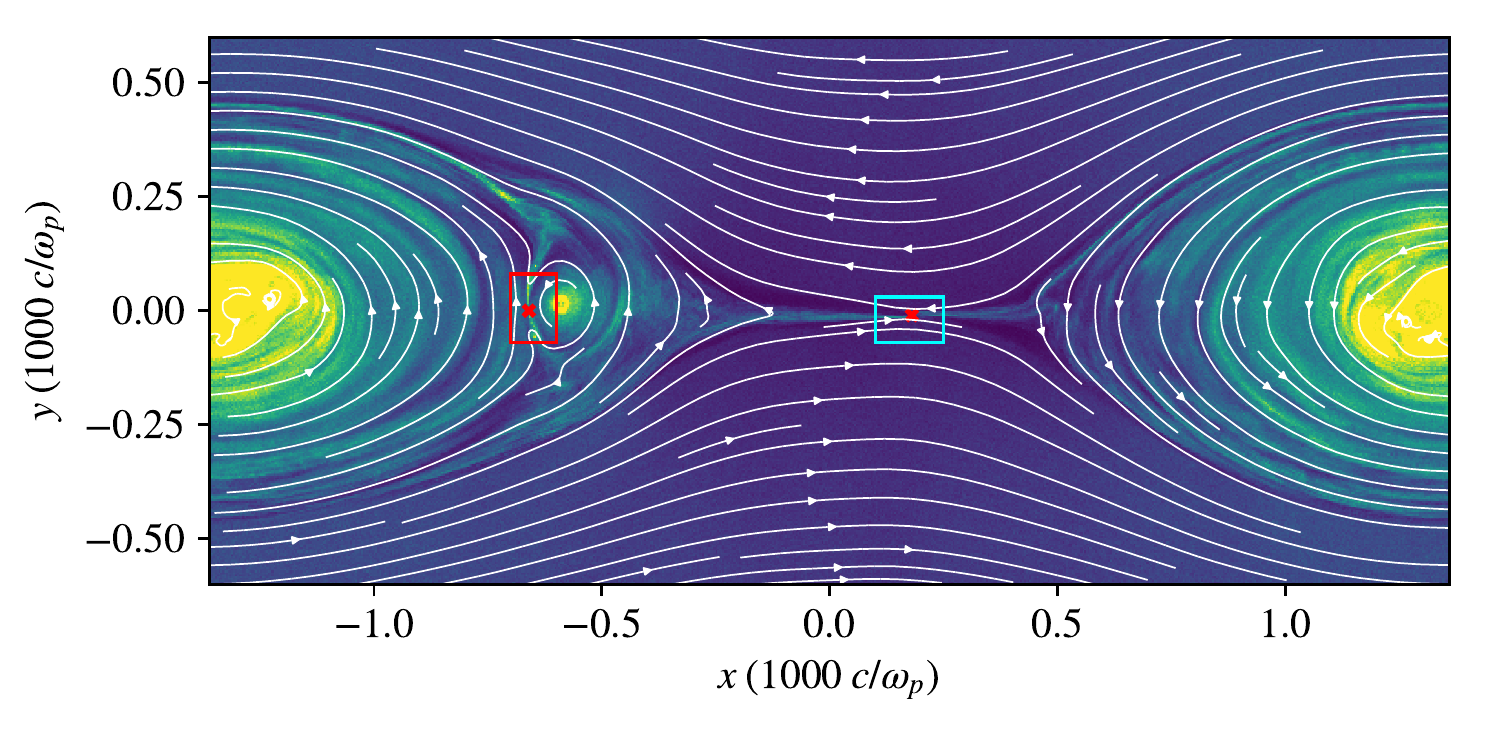}
	\caption{Snapshot of density from a triggered simulation with a guide field of $B_{g}/B_{0}=0.3$ at $t=14700 \; \omega_{p}^{-1}$ (run A0*).  We superimpose streamlines of the in-plane magnetic field and emphasize two regions: the cyan box where reconnection is taking place in the PCS, and the red box, where reconnection is occurring in a MCS.  Note that the sign of $\vec{\nabla}\times \bold{B}$, and hence the sign of the out-of-plane electric field is positive in the horizontal layer (cyan box), and negative in the MCS between plasmoids (red box).  We plot with red crosses the X-points identified from this snapshot.
	}
	\label{blines}
\end{figure*}

In order to achieve different numbers of X-points per unit length for simulations with identical physical parameters (and so, probe the role of X-point acceleration), we use two strategies: (\textit{i}) we let the primary tearing mode evolve spontaneously (hereafter, untriggered runs), or, (\textit{ii}) we trigger reconnection at the center of the domain.  In the former case, we use a sheet thickness of $\Delta=13.33 \; c/\omega_{p}$, where $\omega_{p}$ is the upstream electron plasma frequency,.  In the latter case, we use a thicker sheet of $\Delta=26.66 \; c/\omega_{p}$ to ensure that no primary X-points form other than the one we induce in the center of the domain.  See Table \ref{tab:fit} for a list of all the simulations and their respective physical and numerical parameters we employ in this study.  We now describe in detail these two setups.

\subsection{Fiducial Simulations}
In order to investigate the electron acceleration mechanisms, our first goal is to create distinct realizations of density and field structures.  We find that one numerical and one physical parameter have the highest impact on the structure of the current sheet and can give us a great diversity of current layers.  

The physical parameter we change is the guide field strength.  Specifically, for our particular values of $\sigma$ and $\beta_{i}$, we find that including a guide field of strength $B_{g}/B_{0}=0.3$ suppresses the formation of secondary X-points and plasmoids, while a guide field strength of $B_{g}/B_{0}=0.1$ allows for copious X-point and plasmoid formation throughout the reconnection layer, in analogy to the case with zero guide field.  

The numerical parameter is whether or not reconnection is triggered, which affects the number of primary X-points and primary plasmoids.  All PIC studies of reconnection have to make the choice of whether to trigger reconnection at a specific point(s) in the current sheet or to let it evolve spontaneously.  In a typical triggered setup there is one primary X-point and the Alfv\'en crossing time along the layer is less than the primary tearing growth time (this is guaranteed by choosing a thickness $\Delta$ large enough).  Because of this, a single large magnetic island forms at the boundary when the reconnection fronts collide, and no other primary plasmoids form.  By ``primary'' we are referring to structures (both X-points and plasmoids) that form directly from the initial Harris current sheet: the properties of these primary structures depend on the particular initialization of the current sheet.  In the region in between the reconnection fronts, the secondary tearing mode (when active) self-consistently forms X-points and plasmoids in the center of the domain which are pulled to the edges of the box and ultimately merge with the large boundary island.  The properties of these ``secondary'' X-points and plasmoids depend only on the flow conditions far from the current sheet and have no memory of the initialization of the Harris current sheet.  In our triggered setups, we employ a thick current sheet ($\Delta=26.66 \; c/\omega_{p}$) and remove by hand the pressure\footnote{We reduce by hand the momenta of the hot particles initialized in the sheet, in a region centered in the middle of the layer and extending for 133 $c/\omega_{p}$ along the sheet.} of the hot particles initialized in the current sheet, such that only one primary X-point forms.   
 
 
 In contrast, in an untriggered setup, numerous primary X-points and plasmoids form via the primary tearing mode.  These primary plasmoids then hierarchically merge until there is one large magnetic island.  In this way, the untriggered setup invariably has more primary X-points as well as more large plasmoid mergers than a triggered setup with the same physical parameters \footnote{In between each pair of primary islands, secondary X-points and plasmoids may form, when the secondary tearing mode is active}.  In our untriggered setups, we use a thin current sheet ($\Delta=13.33 \; c/\omega_{p}$), which is sufficiently thin such that there are numerous primary X-points along the layer.

One can make arguments for the applicability of either choice to realistic situations.  For instance, if the astrophysical current sheets of interest are thick, then the growth time of the primary tearing mode is long and reconnection likely will not proceed without some external perturbation.  This situation is likely better described by a triggered setup.  If the current sheet is sufficiently thin such that the timescale of the primary tearing instability is shorter than any relevant dynamical time, then we expect reconnection to spontaneously evolve before the current sheet is dynamically disrupted.   Here, however, we focus less on the concerns of applicability and simply use the choice of whether or not to trigger reconnection as a numerical tool to achieve a diversity of density and electromagnetic field structures to enable us to probe electron acceleration under these varied conditions.

Using these two values for the guide field and the choice of whether or not to trigger reconnection by hand, we create four fiducial simulations.  

\begin{figure*}[t]
	{
		\includegraphics[width=\linewidth]{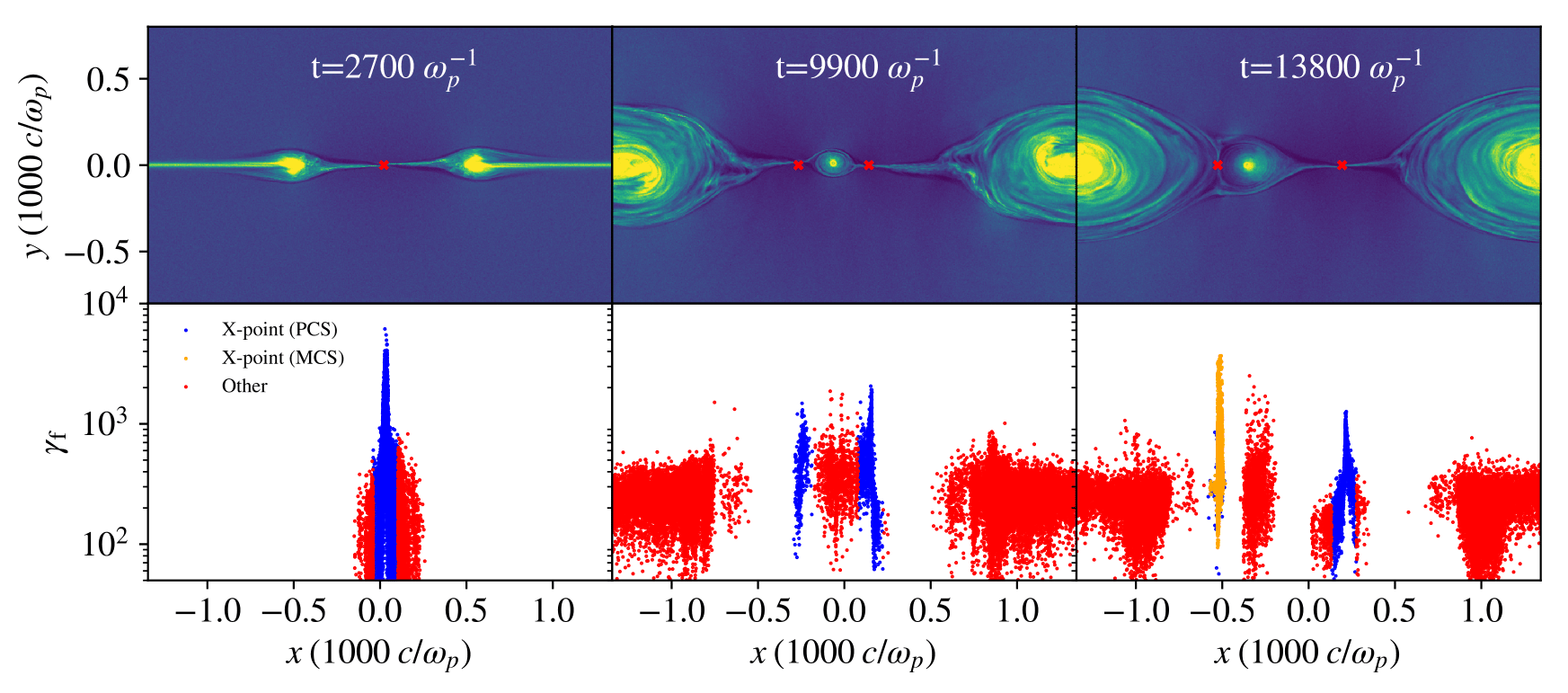}
		\caption{Snapshots at three different times from the triggered simulation with a guide field strength of $B_{g}=0.3B_{0}$ (run A0*).  Each column corresponds to a different time, increasing to the right.  The top panels show snapshots of the density and the locations of X-points are depicted with red crosses.  The bottom panels plot the $x$-position of electrons at the time they first exceed $\sigma_{e}/2$ against their final Lorenz factor $\gamma_{f}$.  If the particle acceleration episode is in the vicinity of an X-point, we color the particle in blue (orange) if the sign of $E_{z}$ positive (negative).  If the particle is not accelerated near an X-point, we color it in red.  We see that the electrons that end up with the highest energies invariably are first accelerated near an X-point.  One secondary plasmoid forms (middle column), and eventually merges into the boundary island (right column), accelerating a number of electrons in the current sheet that forms during the merger (orange points at $x=-500 \; c/\omega_{p}$).
		}
		\label{triggered_bguide_snapshots}}
\end{figure*}
We show snapshots from our four fiducial simulations in Figure \ref{lowbeta_fourdens}, all taken at $t=3600 \; \omega_{p}^{-1}$ or equivalently $t=0.63\; t_{A}$, where $t_{A}=L_{x}/v_{A}$ is the Alfv\'en crossing time of the box, $L_{x}$ is the length of the box in the $x$-direction and $v_{A}$ is the upstream Alfv\'en velocity.  Here, the top row shows snapshots of density from the simulations with $B_{g}/B_{0}=0.3$ while the bottom row shows snapshots from simulations with $B_{g}/B_{0}=0.1$.  The left column shows the triggered simulations while the right column shows the untriggered simulations.  We see that, for the particular $\sigma$ and $\beta_{i}$ we use here, a guide field with strength $B_{g}/B_{0}=0.3$ results in thicker, more stable current sheets that do not fragment via the secondary tearing mode.  This is in stark contrast with lower guide field simulations at the same $\sigma$ and $\beta_{i}$ (bottom row), where the current sheet fragments copiously into secondary X-points and plasmoids.

We show in Table \ref{tab:fit} all of the simulations we employ in this study.  The four fiducial simulations we refer to in the body of this paper are labeled A0*, B0*, C0*, and D0*.  Variations of a fiducial simulation for fixed guide field and triggering mechanism (e.g., varying box size or current sheet thickness) are denoted with the same letter but a different number.  We present the results of these simulations in the appendices.
 \\
 \\

\section{Diagnostics of Electron Acceleration}\label{diagnostics_of_acceleration}
Having set up four varied fiducial simulations, our goal is to investigate how the different structures impact the electron acceleration mechanisms.  This involves (\textit{i}) tracking specific particle properties on-the-fly that serve as diagnostics of their acceleration, (\textit{ii}) identifying X-points from the electromagnetic fields, and (\textit{iii}) devising criteria to classify acceleration episodes to determine the relative importance of different acceleration mechanisms.
\subsection{Tracking Particle Properties on the Fly} \label{diagnostics}
In \citet{ball2018}, we found that high-energy electrons generally experience short episodes of intense acceleration (on timescales of order $\sim 100 \; \omega_{p}^{-1}$).  Because the output cadence is often drastically down-sampled in time as compared to the simulation timestep, it can be difficult to pin down precisely when and where electrons are accelerated.  In addition, particle outputs are often down-sampled in number, i.e., only a small fraction of particles are saved in the output files for analysis.

\begin{figure*}[htp]
	\includegraphics[width=\linewidth]{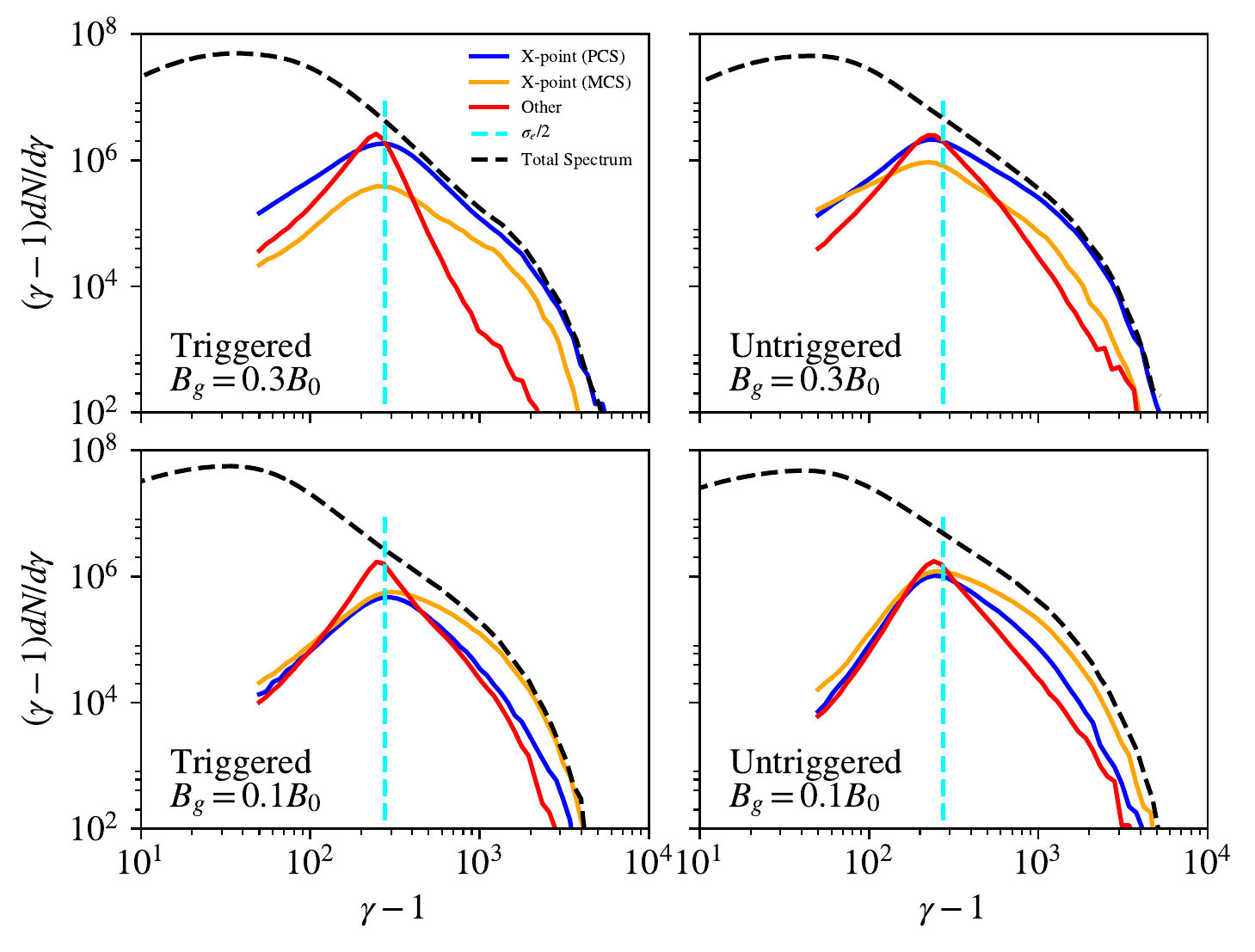}
	\caption{At the final time of our simulations, $t=19,500 \; \omega_{p}^{-1}$, we plot the electron energy spectra from our four fiducial simulations dissected by spatial location of injection.  The blue (orange) line corresponds to particles that were injected near an X-point in the PCS (MCS) and the red line corresponds to particles that were not injected near an X-point.  In the $B_{g}=0.3B_{0}$ case (top row) the highest energy particles are mostly injected near X-points in the PCS, while in the $B_{g}=0.1B_{0}$ case, there are copious plasmoid mergers, and high-energy particle injection is dominated by X-points in MCS.}
	\label{spectra_fourplot}
\end{figure*}

In order to get around these problems, we track four additional properties of particles on-the-fly during the simulation.  At every simulation timestep, we check whether the particle exceeds an energy threshold, $\gamma > \sigma_{e}/2$ for the first time.  Here $\sigma_{e}=B_{0}^{2}/(4 \pi m_{e} n_{e} c^{2})=\sigma_{i} m_{i}/m_{e}$, representing the magnetic energy available per electron.  If this criterion is met, we save the time and the location of the particle at this time, as well as the direction of $E_{z}$ it is experiencing.  We use this particular threshold because we are interested in probing acceleration into the high-energy power-law tail, rather than the thermal peak of the distribution.  As we will show in the electron energy spectra later, an energy of $\gamma=\sigma_{e}/2$ is safely beyond the thermal peak.  Additionally, we keep track of the cumulative work done on particles by the $z$-component of the electric field that is parallel to the local magnetic field.  

\begin{equation}
	W_{||,z} = \frac{1}{m_{e}c^{2}}\int_{0}^{t_{f}}qE_{||,z}v_{z}dt
	\label{EQUATION_wparz}
\end{equation}
where $E_{||}=\bold{E}\cdot\hat{b}$ (and the vector $\boldsymbol{E_{||}}=E_{||}\hat{b}$), and $E_{||,z}$ is the $z$-component of $\bold{E_{||}}$ so $E_{||,z} \equiv E_{||}\hat{b}\cdot \hat{z}$.  Here, $q$ is the charge of the particle, $\hat{b}$ is a unit vector along the local magnetic field, $\bold{E}$ is the electric field, $v_{z}$ is the $z$-component of a particle's velocity, and $t_{f}$ is the time being considered.  This is an especially useful quantity to track because $E_{||,z}$ is the component of non-ideal fields that is associated with X-point acceleration.  In anti-parallel reconnection, the non-ideal reconnection electric field would still be primarily along the $z$-direction.  However, the magnetic fields will be predominantly in the $xy$ plane, so $E_{||}=0$, i.e., $E_{||}$ is not a good proxy of non-ideal electric fields in anti-parallel reconnection.  A small but nonzero guide field component allows us to employ $E_{||}$ as a good diagnostic of non-ideal fields.

These quantities are useful diagnostics for understanding the acceleration mechanisms: the time and location of the particle's first episode of acceleration allow us to explore what structures the particle is interacting with during the time of its first significant energization.  The sign of $E_{z}$ at this time helps distinguish between acceleration in the PCS ($E_{z}>0$) versus in MCS ($E_{z}<0$).  Finally, the work done by the $z$-component of parallel electric fields is useful because it allows us to distinguish work done by non-ideal electric fields associated with reconnecting magnetic fields from other mechanisms such as Fermi-type acceleration from velocity convergences in the inflow and outflow regions which are associated with ideal electric fields.

We check for the $\gamma > \sigma_{e}/2$ condition at every simulation timestep for all electrons and record the time and position when this criterion is satisfied.  We note that this method will only record information about the first acceleration episode that an electron experiences.  This first episode, however, is critical to promoting electrons to relativistic energies, which allows them to sample large-scale velocity differences and become further energized through Fermi-type processes.

In order to explore acceleration after the electron's promotion out of the cold $\gamma\approx1$ population to highly relativistic energies, we also follow a sample of electron trajectories and explore the contributions of $E_{||,z}$ to the acceleration of a typical high-energy electron.  In general, we find that the highest energy electrons are almost always first accelerated by $E_{||,z}$ at an X-point, and then are further accelerated by a combination of $E_{||}$ in current sheets during plasmoid mergers and $\boldsymbol{E}_{\bot}=\boldsymbol{E} \times \hat{b}$ associated with the interaction of outflows with plasmoids or with the turbulent motions within plasmoids.

\subsection{X-point Identification} \label{xpoint_id}
In order to test the association of electron energization episodes with X-points, we first identify X-points from the fields.  \citet{haggerty2017} recently studied the statistics of X-points in 2.5D turbulence via PIC simulations and explored methods to robustly identify X-points.  In a 2.5D setup such as ours, X-points correspond to saddle points in the $z$-component of the magnetic vector potential, $A_{z}$.  Following \citet{haggerty2017}, we first apply a Gaussian filter with a width of $\sim 4 \; c/\omega_{p}$ to the $z$-component of the magnetic vector potential, $A_{z}$.  We then identify critical points where $\partial A_{z}/\partial x=\partial A_{z}/\partial y=0$.  In order to distinguish between local minima, maxima, and saddle points, we calculate the matrix of second derivatives (or, Hessian matrix), 
$$H_{ij}=\frac{\partial^{2} A_{z}}{\partial x_{i} \partial x_{j}}.$$  If the eigenvalues of this matrix are of opposite sign, then this is a saddle point and we identify it as an X-point.  We apply one more constraint to the identification of X-points, by requiring that they reside in cells where the density of hot particles initialized in the current sheet is $<N_{ppc0}$.  This ensures that we exclude saddle points created by noise-level pinches in the initial unperturbed current layer, which would not necessarily develop into active X-points.

We show in Figure \ref{blines} a snapshot from a triggered simulation with a guide field strength of $0.3 B_{0}$ at $t=14700 \; \omega_{p}^{-1}$ (run A0*), where a secondary plasmoid merges into the boundary island at the end of the simulation.  We plot the locations of X-points identified with the method described above with red crosses.  We see that we are able to identify X-points not only in the initial horizontal current sheet (PCS), but also X-points generated in the current sheets at the interface of merging plasmoids (MCS).

\begin{figure*}
	\includegraphics[width=\linewidth]{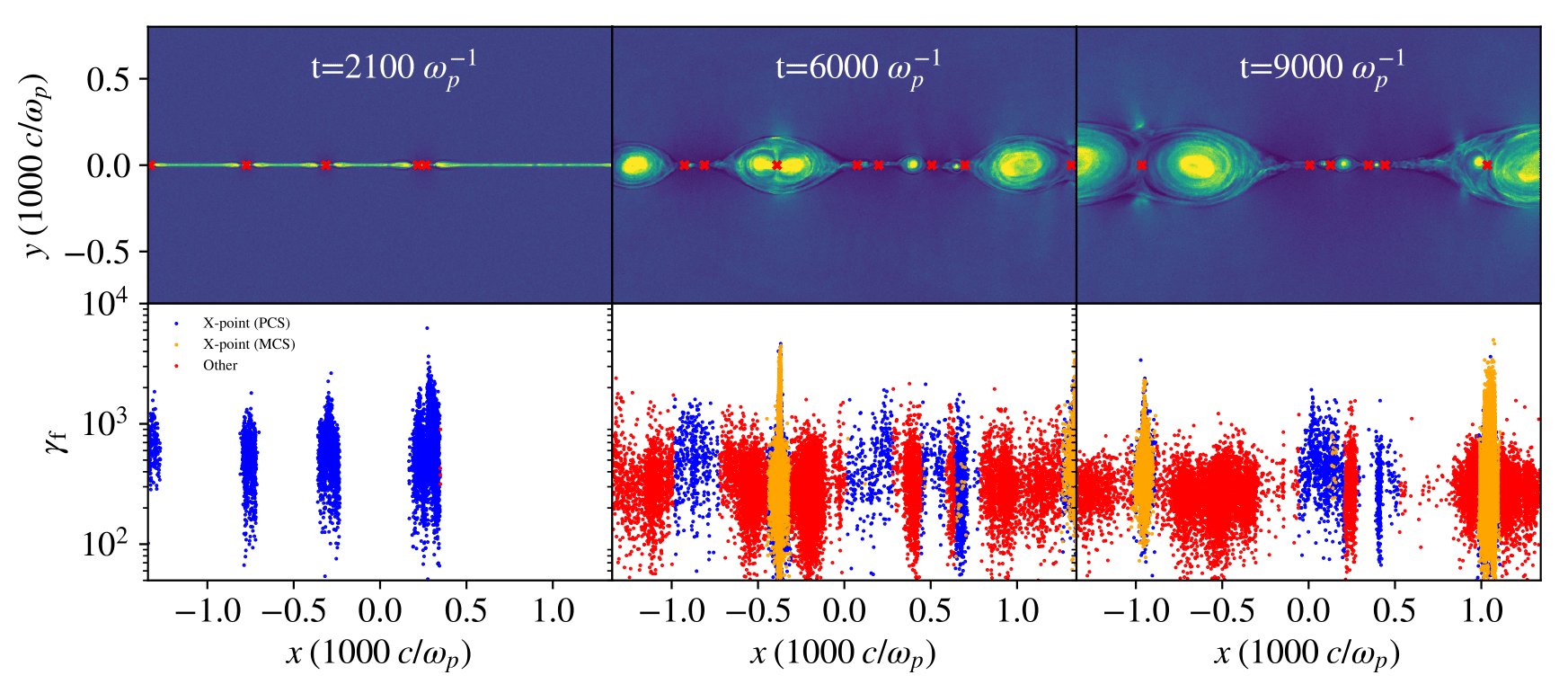}
	\caption{Snapshots at three different times from the untriggered simulation with a guide field strength of $B_{g}=0.1B_{0}$.  We see that both the primary and secondary tearing mode result in copious X-point and plasmoid formation and that the prevalence of these structures results in enhanced efficiency of electron acceleration.
			}
	\label{untriggered_bguide1_snapshots}
\end{figure*}

\subsection{Spatial Locations of Electron Injection} \label{acc_mec_def}
Our goal is to determine the relative importance of X-points in the PCS and MCS in accelerating electrons to relativistic energies.  In order to do this, we use a few simple criteria to distinguish between electrons that are accelerated near an X-point in the initial horizontal current layer from those at the interface of two merging plasmoids.
If a particle is accelerated at an X-point in the PCS, then we expect the particle to experience a sudden non-ideal out-of-plane (in the $+\hat{z}$ direction) electric field during its first interaction with the current sheet.  Conversely, a particle that is accelerated in a MCS will experience a similar episode of acceleration, but with the opposite sign of electric field (in the $-\hat{z}$ direction).

We illustrate this in Figure \ref{blines}, where we show in-plane magnetic field lines superimposed on a snapshot of density from the triggered simulation with a guide field strength of $B_{g}=0.3B_{0}$ at $t=14700 \; \omega_{p}^{-1}$.  We see a typical X-point in the PCS, highlighted by a cyan box.  Note that $\vec{\nabla} \times \bold{B}$ in this region is in the $+\hat{z}$ direction.  At $x\approx-600 \; c/\omega_{p}$ we see a secondary plasmoid merging into the large boundary island, highlighted by a red box.  A vertical current sheet forms, with $\vec{\nabla} \times \bold{B}$ and corresponding electric field in the $-\hat{z}$ direction.

To find all the electrons at a given time that are accelerated by X-points, we first identify X-points from the fields as described above in Section \ref{xpoint_id}.  We note that in this part of our study, we are sensitive to the output temporal cadence: we cannot output the fluid structure and identify X-points at every timestep.  We output the fluid structure once every  $150 \; \omega_{p}^{-1}$, but have tested shorter output cadences and found that our conclusions are not sensitive to this choice. Generally, after an electron is accelerated at an X-point, it either enters a plasmoid or the unstructured outflow and moves away from the X-point.  The outflow moves at $\sim v_{A}$, while plasmoids generally move slower than this; the bulk flow of plasma away from X-points is no faster than $\sim v_{A}$.  Because of that, we look for all electrons that have acceleration episodes that are ``Alfv\'enically connected'' to an X-point, i.e., we require that 

\begin{equation}\label{alfven_connection}
\left\lvert\frac{x_{\textrm{thresh}} - x_{\rm{xpoint}}}{t_{\textrm{thresh}}-t_{\rm{xpoint}}}\right\rvert \le v_{A}.
\end{equation}
Here, $t_{\textrm{thresh}}$ and $x_{\textrm{thresh}}$ are the time and location, respectively, of the particle when it first satisfies $\gamma \geq \sigma_{e}/2$. $x_{\rm{xpoint}}$ is the location of the nearest X-point, $t_{\rm{xpoint}}$ is the nearest output time to $t_{\rm{thresh}}$ from which we identify X-points, and $v_{A}$ is the Alfv\`en velocity.  If an electron's acceleration episode satisfies equation \ref{alfven_connection}, then we classify the particle as being injected near an X-point\footnote{Electrons can move away from an X-point at close to the speed of light, but we use Alfv\'enic causal connection as an even more constraining criterion for relating acceleration episodes to X-points.}.

In order to further distinguish between injection at an X-point in the PCS versus a MCS, we use the fact that the electric field will have opposite directions in the two cases.  Therefore, if the $E_{z}$ field is negative at the time of a particle's injection and the particle is Alfv\'enically connected to an X-point, we identify the acceleration episode as being due to an X-point generated in a merger.

If the particle is not accelerated near an X-point, we classify the acceleration episode as ``other''.  We find that these uncategorized episodes generally produce the majority of lower-energy electrons and they are often associated with plasmoid motion, contraction, or the interaction of an outflow with a plasmoid.  However, as we show below, most of the high-energy electrons experience their first episode of acceleration in the vicinity of an X-point.


We note that this criterion is merely based on spatial proximity of an injected particle with an X-point, but it remains agnostic about the nature of the particle acceleration mechanism. In particular, this criterion will flag as “electrons connected to X-points” both particles that are accelerated by the non-ideal electric field at X-points, as well as electrons (if any) that are energized by ideal fields in the outflow regions near X-points. In Section \ref{E_comps} and \ref{test_prtls}, we take a closer look at the role of non-ideal fields for electron injection.

\section{The Roles of X-points  in Electron Injection}\label{role_of_xpoint}
In order to assess the relative importance of X-points in the PCS and MCS in shaping the overall electron energy spectra, we examine the spectra from the different injection locations.  That is, at each output time, we identify the location of X-points as described in Section \ref{xpoint_id}.  We then associate all the electrons with an injection location (either X-points in the PCS or MCS, or ``other'') as described in Section \ref{acc_mec_def}.  We then construct energy spectra from these different components and assess their relative importance.  To illustrate this process, we show in Figures \ref{triggered_bguide_snapshots} and \ref{untriggered_bguide1_snapshots} snapshots from the classification scheme for the triggered $B_{g}=0.3B_{0}$ and untriggered $B_{g}=0.1B_{0}$ cases.  We show these particular cases because they show the least and most secondary structures, respectively.  We show in Figure \ref{spectra_fourplot}, however, the dissected spectra for all four fiducial simulations.

In the top panels of  Figure \ref{triggered_bguide_snapshots} we show the density at three snapshots in time from the triggered simulation with a guide field strength of $B_{g}=0.3B_{0}$.  The locations of X-points at that time are depicted with red crosses.  In the bottom panels, we plot the injection location (the location of the electron when it first satisfies $\gamma > \sigma_{e}/2$), $x_{\rm{thresh}}$, versus the final energy of electrons, $\gamma_{\rm{f}}$, measured at the end of our simulation at $t=19500 \; \omega_{p}^{-1}$.  If the electron is accelerated near an X-point in the PCS (MCS), we color it in blue (orange), otherwise, we color it in red.  We see that at an early time (left column), there is a single primary X-point accelerating electrons.  As reconnection proceeds, a secondary plasmoid begins to develop in the middle of the domain (middle column), and there are two corresponding secondary X-points on either side.  These secondary X-points also accelerate electrons, but not as prolifically as the initial primary X-point.  Eventually, the plasmoid is pulled towards the left edge of the domain  and merges with the large boundary island (right column).  A current sheet forms between the two merging plasmoids, serving as another site of acceleration, with the expected flip in $E_{z}$ polarity as compared to the X-points in the initial current sheet.  


We show the spectra decomposed by injection location for our four fiducial simulations in Figure \ref{spectra_fourplot}.  We see that in the triggered $B_{g}=0.3B_{0}$ case we examined in Figure \ref{triggered_bguide_snapshots}, corresponding to the top left panel of Figure \ref{spectra_fourplot}, the majority of high-energy electrons (defined as $\gamma \geq \sigma_{e}/2 \approx 300$) are injected near X-points, and in particular, the primary X-point.  This is because the guide field suppresses secondary X-point and plasmoid formation, resulting in only one secondary plasmoid (an so, one merger) during the simulation.  We note that because acceleration is so localized in this case, in the limit of very large domains, the acceleration region will comprise a vanishing fraction of the total domain and we expect the acceleration efficiency to be negligible (see Appendix \ref{box_size} for the dependence of our results on domain size.).  

\begin{figure*}[t]
	\includegraphics[width=\linewidth]{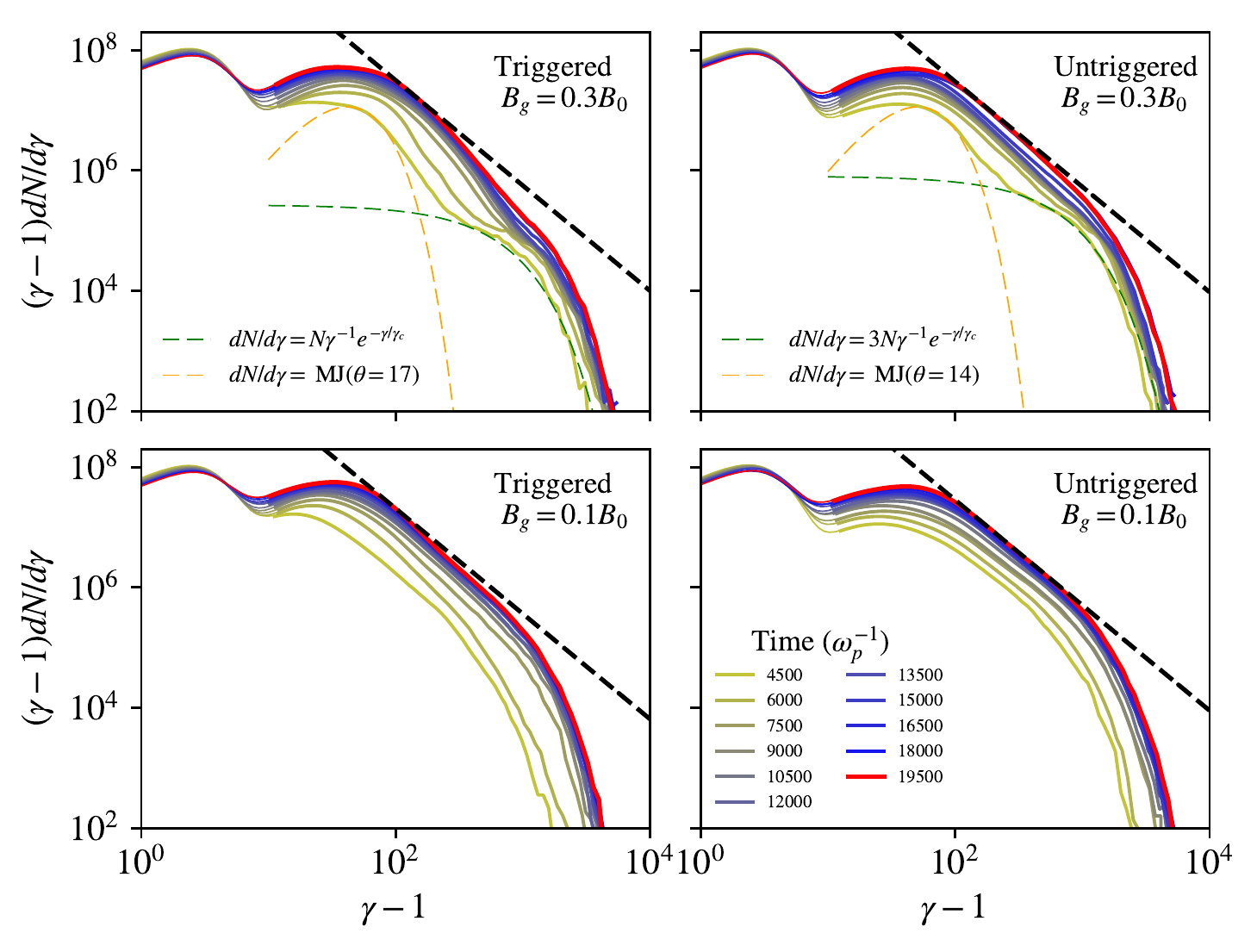}
	\caption{Time evolution of the overall electron energy spectra from our four simulations.  The thick component of each line shows the spectrum taken only in the reconnection region while the thin component corresponds to the colder upstream plasma.  Yellow lines correspond to early times and blue lines correspond to later times.  The red line in each panel depicts the spectrum at the latest time in that simulation.  A power-law distribution with $p=2.7$ is depicted with a dashed black line in all panels for reference, normalized to lie tangent to the spectra near their respective thermal peaks in the post-reconnection region.  In the $B_{g}=0.3B_{0}$ cases, we emphasize the two-component nature of the spectra at early times by plotting a Maxwell-J\"{u}ttner distribution with an appropriate temperature and normalization to match the low-energy bump in the reconnection region as well as a power-law with an index of 1 and an exponential cutoff at $\gamma=450$. 		
		Only the triggered $B_{g}=0.3B_{0}$ case shows a significantly softer spectrum due to a suppressed number of X-points.}
	\label{2x2_timespec}
\end{figure*}
We show the same decomposition, but for an untriggered simulation, with the same guide field of $B_{g}=0.3B_{0}$, in the top right panel of Figure \ref{spectra_fourplot}.  A snapshot of the density structure of this simulation can be seen for reference in the top right panel of Figure \ref{lowbeta_fourdens}.   We see that X-points in the PCS continue to dominate the injection of high-energy electrons, as in the triggered case with the same guide field strength.  In summary, we find that for guide fields of $B_{g}/B_{0}=0.3$, most high-energy electrons are injected at primary X-points in the PCS. Secondary X-points and island mergers play a sub-dominant role, since the guide field suppresses the secondary tearing mode.

In Figure \ref{untriggered_bguide1_snapshots} we show three snapshots from the untriggered simulation with $B_{g}=0.1B_{0}$ (run D0*).  We show this simulation because it is the one with the largest number of X-points and plasmoids due to the fact that the secondary tearing mode is active, as well as it being an untriggered simulation (so, with numerous primary X-points and primary plasmoids).  This guarantees that there will be numerous mergers of varying plasmoid sizes, from small secondary plasmoids merging with one another, to the larger primary plasmoids hierarchically merging until the system is dominated by a single large plasmoid.  At early times (left column), numerous primary X-points form and begin accelerating electrons at these specific locations.  In the middle column, we see multiple large primary plasmoids, two of which are merging at $x \approxeq -400 \; c/\omega_{p}$, while the other two primary plasmoids are just beginning to merge across the periodic boundary, as evidenced by the orange points at the rightmost edge of the middle column.  In addition, we see the formation and subsequent merging of secondary plasmoids throughout the reconnection layer.  The X-points that occur frequently in the primary layer continuously accelerate electrons (e.g., at $x \sim 0 \; c/\omega_{p}$ in the rightmost panel), while X-points in the MCS also contribute considerably to high energy electron acceleration (e.g., at $x=1000 \; c/\omega_{p}$ in the rightmost panel). 

We show in the bottom right panel of Figure \ref{spectra_fourplot} the spectra of electrons from the simulation shown in Figure \ref{untriggered_bguide1_snapshots}.  We see that, due to the large number of plasmoid mergers that occur all throughout the domain, the majority of the high-energy electrons are first energized in MCS.  X-points in the PCS still play a considerable role, but are not dominant like they were in the case where the secondary tearing mode was suppressed, at higher guide fields.  It is not clear from this analysis where the majority of the ``other'' particles come from: it is possible they interact with X-points as they enter the current sheet for the first time but do not exceed $\sigma_{e}/2$ (so, they do not pass our energy threshold) or are simply engulfed into a plasmoid from the upstream.  In either case, these electrons typically do not reach energies as high as electrons that are first accelerated beyond $\gamma=\sigma_{e}/2$ in the vicinity of an X-point: the vast majority of electrons that exceed $\gamma  > \sigma_{e}/2$ are first accelerated near an X-point.

We show in the bottom left panel of Figure \ref{spectra_fourplot} the spectra decomposed by acceleration mechanisms from our triggered $B_{g}=0.1B_{0}$ simulation.  A snapshot of the density structure of this simulation can be seen for reference in the lower left panel of Figure \ref{lowbeta_fourdens}.  We see that X-points in MCS dominate the overall spectrum due to onset of the secondary tearing mode, similarly to the untriggered case with the same guide field strength.  In summary, when the guide field is weak (in analogy to the purely anti-parallel case), the secondary tearing mode is active, resulting in numerous plasmoid mergers and hence efficient electron injection at MCS.

The diversity of injection locations and mechanisms for high-energy electrons is also imprinted in the evolution of the electron energy spectrum.  We show in Figure \ref{2x2_timespec} the time evolution of the electron energy spectra from each of our four simulations.  We show the spectra from the ``reconnection region'' (i.e., excluding the upstream plasma; see \citealt{ball2018} for more details on how this region is selected) with thick lines, and spectra including the upstream plasma are shown with thin lines.  Spectra from different snapshots in time are depicted with different colors, with yellow corresponding to early times and blue corresponding to late times.  The red line shows the spectrum from the last time in that simulation.  For reference, we plot the same power-law distribution of $p=2.7$, where $p=-dN/d\log(\gamma-1)$, in each panel with a black dashed line, corresponding to the best fit power-law\footnote{We use the spectral index as a proxy for the hardness of the non-thermal tail and so the efficiency of electron injection.  In reality the electron spectrum (at least for strong guide fields) can be divided into two components with different normalizations, where the lower energy bump is from interactions with the outflow, while the higher energy bump is from acceleration at X-points.  The relative normalization of these two bumps sets the hardness, and hence, the power-law index.} for the hardest spectrum among these four simulations, the untriggered case with $B_{g}=0.1B_{0}$ (lower right).  We see the triggered $B_{g}=0.1B_{0}$ case and untriggered $B_{g}=0.3B_{0}$ case (lower left and upper right, respectively) have spectra that are almost as hard as the untriggered $B_{g}=0.1B_{0}$ simulation, while the triggered $B_{g}=0.3B_{0}$ case has a significantly softer spectrum due to the difference in the total number of X-points.  We see these differences become even more pronounced for larger domains (see Appendix \ref{box_size}).  

By examining the time evolution of the spectra in Figure \ref{2x2_timespec}, we can identify clear connections to the injection mechanisms.  In the cases with a guide field of $B_{g}=0.3B_{0}$ (top row), there is clear evidence of two distinct populations of electrons at early times: a thermal bump at $\gamma\approx 80$ and a distinct bump in the spectrum at higher energies, extending to $\gamma \sim 1000$.  As we have seen in Figure \ref{triggered_bguide_snapshots} and will discuss further in Section \ref{E_comps}, the thermal peak primarily consists of electrons that are injected in locations far from X-points, while the high-energy component consists of electrons that interact with the strong non-ideal field in the vicinity of an X-point.  In order to elucidate the two-component nature of the distribution in the $B_{g}=0.3B_{0}$ cases, we plot a Maxwell-J\"uttner (MJ) distribution (dashed orange) with an appropriate temperature and normalization to match the lower-energy peak in the reconnection region, and also a power-law with an index of $p=1$ with an exponential cutoff of $\gamma_{c}=450$ (dashed green).  We note that the normalization of the power-law component is three times higher in the untriggered case (top right) as compared to the triggered case (top left), which arises naturally because there are three primary X-points in the former as compared to the single primary X-point in the latter.  The higher normalization of the high-energy bump results in a harder spectrum in the untriggered case and is a direct consequence of there being more primary X-points per unit length in the untriggered setup.

In contrast to the cases with a guide field of $B_{g}=0.3B_{0}$, the lower guide field counterparts very quickly evolve into a smoother power-law like distribution.  This is most likely because there are a large number of secondary X-points and plasmoids in the current layer which all accelerate electrons to slightly different energies, smoothly filling in the spectrum in the non-thermal tail of the distribution.

\section{Investigating the Role of non-ideal Electric Fields}\label{E_comps}
In the previous sections we explored where electrons are first accelerated and found that X-points appear to be important structures in regulating the production of a non-thermal electron spectrum.  In this section, we explore the role of the non-ideal electric fields in injecting electrons into a high-energy distribution, as well as the dominant channel of acceleration, i.e., what mechanism is responsible for the majority of energy gain for the highest energy electrons. 

\begin{figure*}[htp] 
	\includegraphics[width=\linewidth]{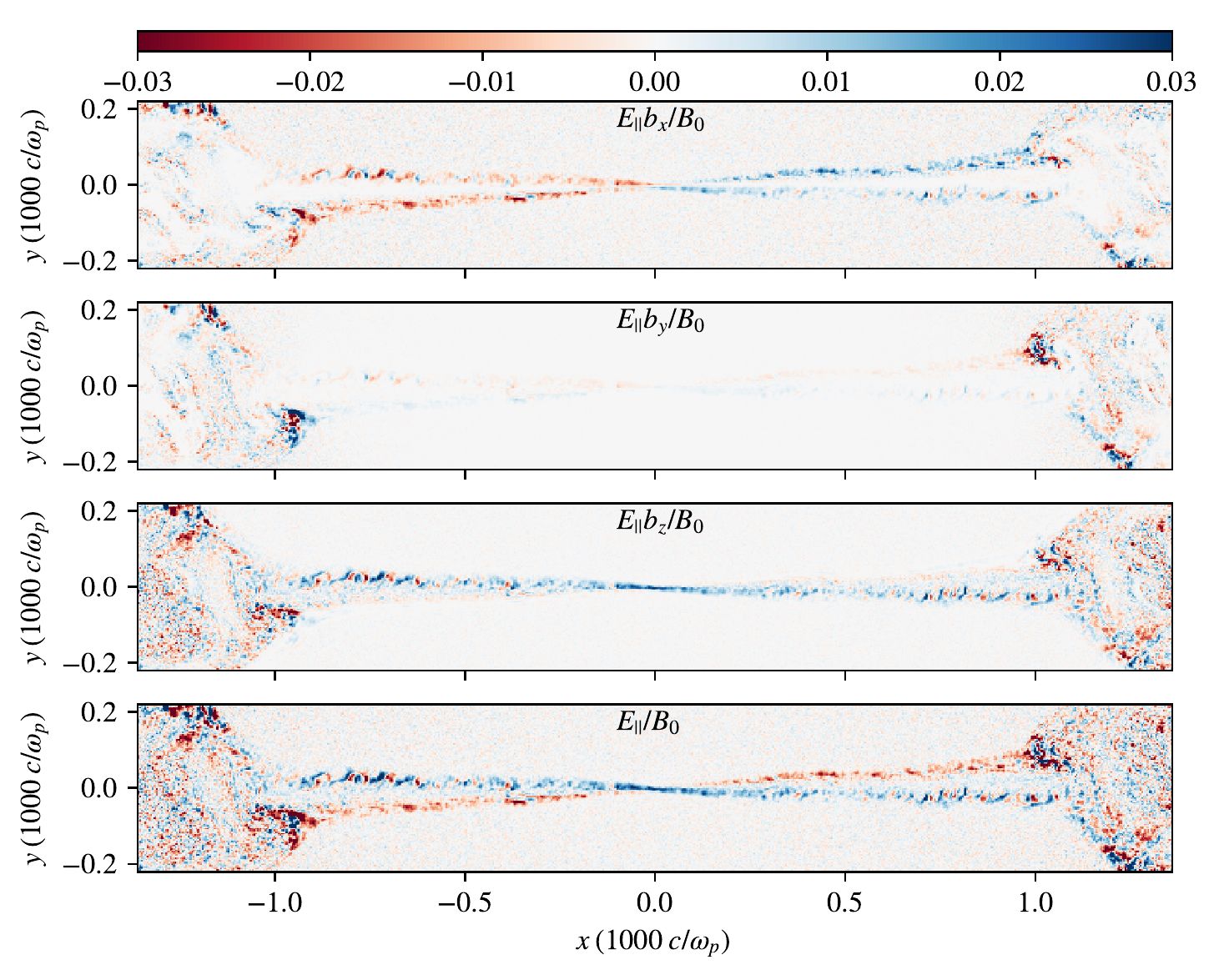}
	\caption{2d structure of the parallel electric field decomposed into its Cartesian components (top three) from the triggered simulation with $B_{g}=0.3B_{0}$.  The bottom panel shows the total $\bold{E} \cdot \hat{b}$.  A significant parallel electric field persists in the edges of the outflow, largely from the $x$-component of $E_{||}$ (first panel).  The $y$-component is negligible except in small localized regions where the dense parts of the outflow are impacting the magnetic island, generating turbulent motions in the large boundary island (i.e., at $x\approx \pm 1000 \; c/\omega_{p}$, $y\approx \pm 100 \; c/\omega_{p}$ in the second from top panel).  The parallel electric field near the X-point is dominated by the $z$-component (third panel from top).}
	\label{edotb_comps}
	\end{figure*}

Reconnection requires non-ideal electric fields.  At X-points, the non-ideal fields are dominated by the $z$-component.  In the presence of a guide field, non-ideal effects can be captured by $E_{||}=\bold{E}\cdot\hat{b}$.  At an X-point, the dominant component of $\boldsymbol{E_{||}}$ is in the $z$-direction.  In order to illustrate this, we show the structure of $\boldsymbol{E_{||}}$ and its various components from our triggered simulation with $B_{g}=0.3B_{0}$ in Figure \ref{edotb_comps}.  The parallel electric field near the X-point is dominated by the $z$-component of $\bold{E_{||}}$, as expected.  In the outflows, however, there is a significant component of $\boldsymbol{E_{||}}$ in the $x$-direction (at the boundary between the outflow and the upstream).   

\begin{figure}[htp] 
	\includegraphics[width=\linewidth]{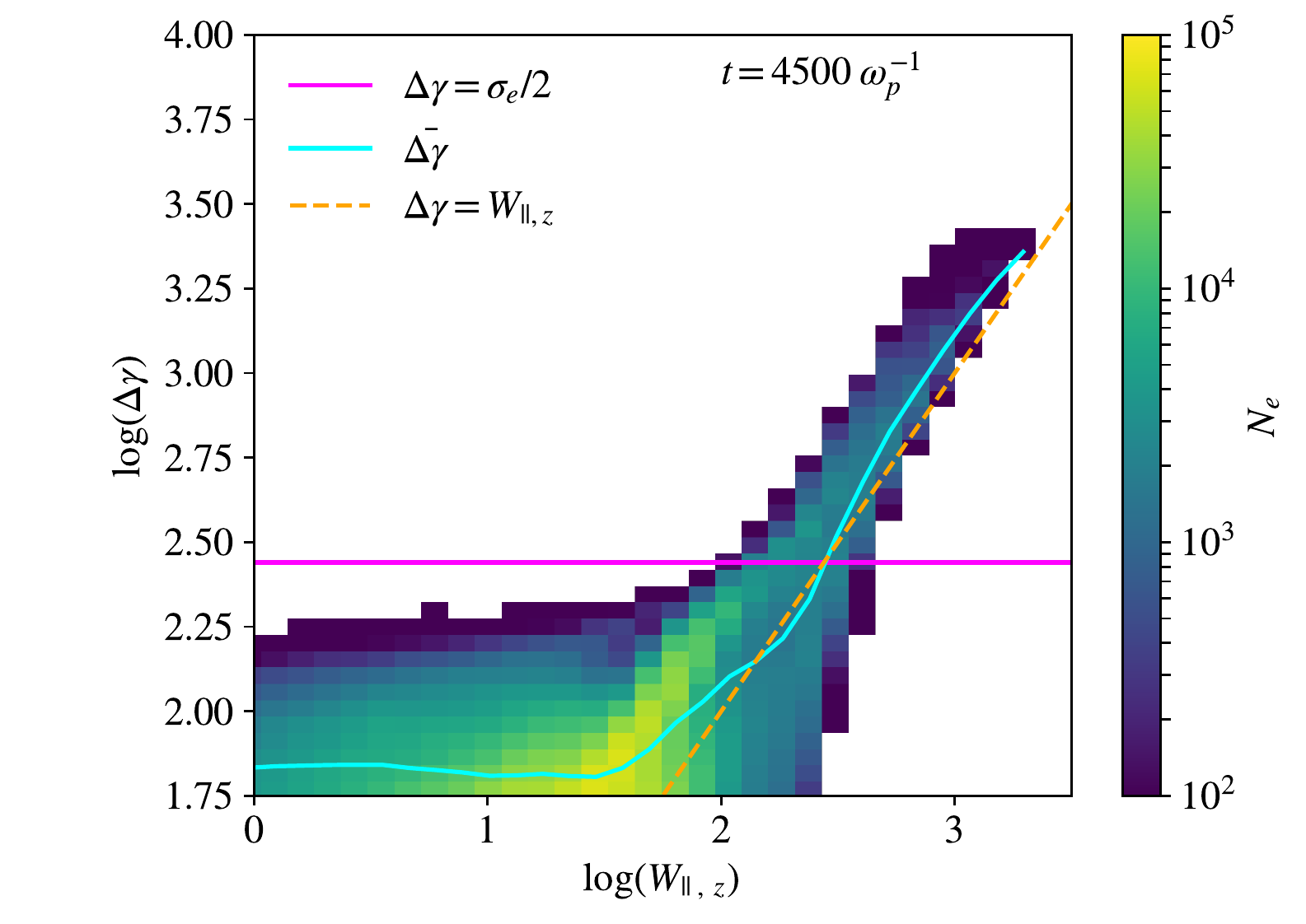}
	\newline
	\includegraphics[width=\linewidth]{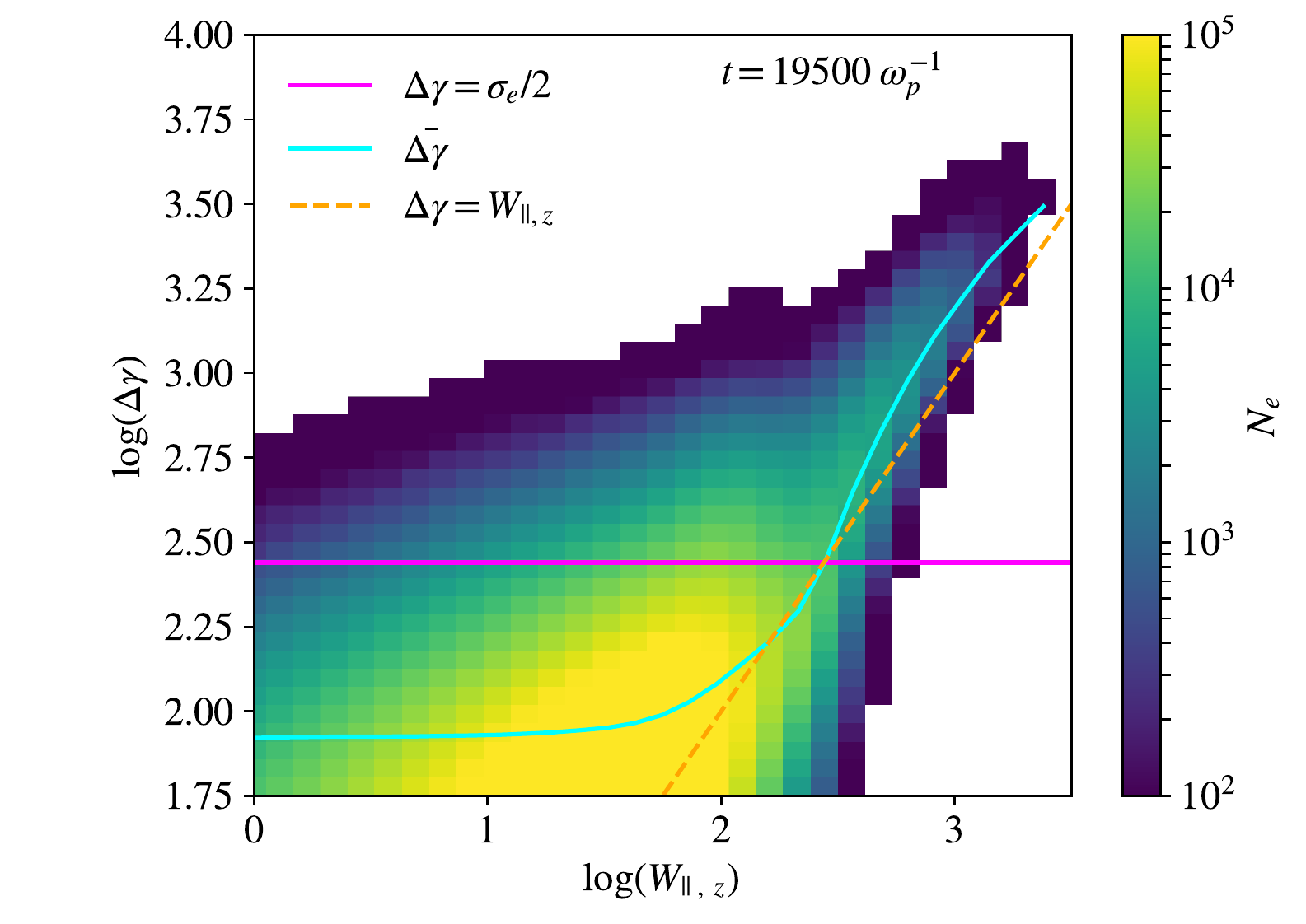}
	
	\caption{2d histograms of the $z$-component of the total change in Lorentz factor versus $W_{||,z}$  taken at two different times (4,500 and 19,500 $\omega_{p}^{-1}$ or, equivalently, 0.78 and 3.37 $t_{A}$) from the triggered $B_{g}=0.3B_{0}$ simulation.  The dashed orange line depicts $\Delta \gamma = W_{||,z}$, the cyan line depicts the average change in Lorentz factor at a given $W_{||,z}$, and the magenta line shows $\Delta \gamma = \sigma_{e}/2$.  The top panel corresponds to a relatively early time in the simulations when the reconnection fronts have not interacted across the boundary, as shown in the bottom left panel of Figure \ref{lowbeta_fourdens}.  The bottom panel corresponds to the final time in the simulation, where a large magnetic island has formed at the boundary and a single secondary plasmoid has merged into it.}
	\label{wpar_hist_earlytime}
\end{figure}

Now that we have an understanding of where in the reconnection region different components of $\boldsymbol{E_{||}}$ are present, we aim to understand their effect on individual electrons and ultimately pin down which features in the spectrum are influenced by these different components.  To this end, we track in our simulation the total work done on each particle by the parallel component of the electric field in the $z$-direction, $W_{||,z}$ (see Equation \ref{EQUATION_wparz}).  We show in Figure \ref{wpar_hist_earlytime} a 2d histogram of the $z$-component of parallel work versus the total change in Lorentz factor, $\Delta \gamma = \gamma - \gamma_{0}$, where $\gamma_{0}$ is the electron's initial Lorentz factor, from our triggered simulation with $B_{g}=0.3B_{0}$ at both an early time before the formation of any magnetic islands (top) and a later time (bottom) when the boundary island has formed and a single plasmoid has formed and merged into it (see Figure \ref{triggered_bguide_snapshots}).  For reference, we show the $\Delta \gamma=\sigma_{e}/2$ line in magenta and the $\Delta \gamma=W_{||,z}$ relation with a dashed orange line.  We also show the average $\Delta \gamma$ at a given $W_{||,z}$ in cyan.  

At early times, the strong correlation between $\Delta \gamma$ and $W_{||,z}$ indicates that the parallel electric field in the $z$-direction at the primary X-point is responsible for accelerating all electrons above $\Delta \gamma \sim 200$.  At later times the dispersion away from the $\Delta \gamma = W_{||,z}$ line points to processes not associated with $E_{||,z}$ that further increase the electron energy.  This follows from the dynamical evolution of the layer: the only available mechanisms to energize electrons to high Lorentz factors at early times are from the non-ideal field at the primary the X-point.  At later times electrons can be energized through the interaction of the outflow with the magnetic island, in the turbulence generated when the reconnection fronts interact across the boundary, in the vicinity of a plasmoid merger, or in a contracting magnetic island.  Although in our case, $E_{||,z}$ controls the injection of electrons into a hard distribution, the majority of energy gain at late times can come from ideal fields (e.g., \citealt{guo2019}).  


\begin{figure}[t]
\includegraphics[width=\linewidth]{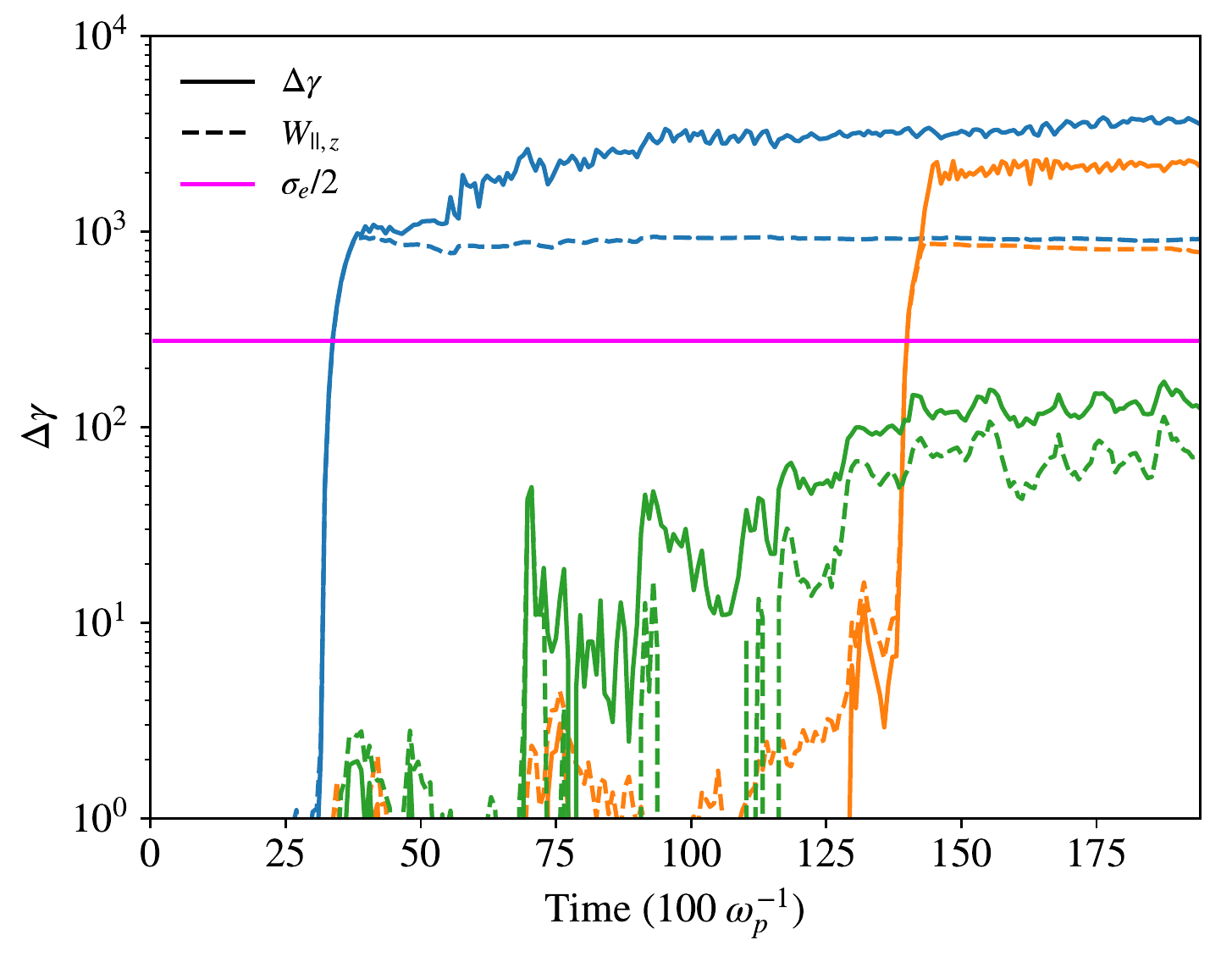}
\caption{Evolution of Lorentz factor ($\Delta \gamma = \gamma - \gamma_{0}$, where $\gamma_{0}$ is the electron's Lorentz factor at $t=0$) and the work done by the $z$-component of parallel electric fields, $W_{||,z}$, of three representative electrons from the triggered simulation with $B_{g}=0.3B_{0}$.  The solid lines represent the electron's Lorentz factor and the dashed lines depict $W_{||,z}$.  }
\label{EdotV}
\end{figure}

We test this understanding further by tracking individual electrons.  In Figure \ref{EdotV} we show the evolution of a few representative electrons' Lorentz factor as well as the work done by the $z$-component of the parallel electric field from the triggered simulation with $B_{g}=0.3B_{0}$.  Each color refers to an individual particle, with the solid line corresponding to the particle's total change in Lorentz factor, and the dashed line representing $W_{||,z}$.  The highest-energy particle (blue line) is typical of the highest energy electrons in the system: it is first accelerated well beyond $\sigma_{e}/2$ in an interaction with the primary X-point, where it is rapidly accelerated to $\gamma \sim 1000$ by $E_{||,z}$.  After this, it continues to gain energy through channels not associated with $E_{||,z}$.  While the X-point phase was critical to this electron's acceleration, we emphasize that the overall energy budget is dominated by the more gradual late-stage acceleration: the X-point phase accelerated the electron to $\gamma \approx 1000$, but the electron is further energized via other processes (i.e., not by $E_{||,z}$) to $\gamma \approx 4000$. The orange particle's trajectory is different: it is first energized to $\gamma \sim 10$ when it enters the secondary plasmoid at an X-point (at around $t=13000 \omega_{p}^{-1}$).  As the secondary plasmoid merges into the boundary island, this particle interacts with the current sheet in between the two and experiences a large $E_{||,z}$ at an X-point in the MCS, which has the effect of increasing the work done by the non-ideal electric field, accelerating the electron to $\gamma \sim 1000$.  The green particle, in contrast, never experiences a particularly strong $E_{||,z}$, and hence its energy remains in the thermal peak of the spectrum.  

In this particular case where the secondary tearing mode is suppressed ($B_{g}=0.3B_{0}$), we see strong evidence for two populations of electrons at early times (see Figure \ref{2x2_timespec}) which we can relate to the 2d histograms of Figure \ref{wpar_hist_earlytime}.  Electrons effectively have two energization paths: they are either heated in the outflows to thermal Lorentz factors of $\sim100$, or accelerated in the vicinity of the primary X-point up to $\gamma \sim 1000$.  The hardness of the total distribution, then, is set by the relative normalization of these two populations: the energy spectrum connects a high-normalization low-energy thermal core out to the low-normalization high-energy component.  In general, X-points comprise a small fraction of the total length of the sheet, which is why the high-energy component of the spectrum has fewer electrons associated with it.  The relative probability that an electron interacts with an X-point as opposed to the outflow and hence be accelerated as opposed to heated depends on the number of X-points per length of the current sheet: if there are more X-points in a current sheet of a given length, we expect the slope to be harder because it is more likely that any given particle will interact with an X-point.  

We can see evidence for this interplay in the top row of Figure \ref{2x2_timespec}: the untriggered case (top right) has a high energy component with a normalization almost exactly three times higher than that in the triggered case (top left), which naturally occurs because there are three primary X-points in the untriggered case as compared to the single primary X-point in the untriggered case.  We test this hypothesis further in Appendices \ref{box_size} and \ref{thickness} where we vary the number of X-points per unit length by varying both the box size in a triggered setup, and the initial sheet thickness in an untriggered setup.  We ultimately find that the number of X-points per unit length is indeed correlated with the hardness of the spectrum: the more X-points per unit length in the current sheet there are, the more high-energy electrons are accelerated and the harder the spectrum is.

\begin{figure}[htp] 
	\includegraphics[width=\linewidth]{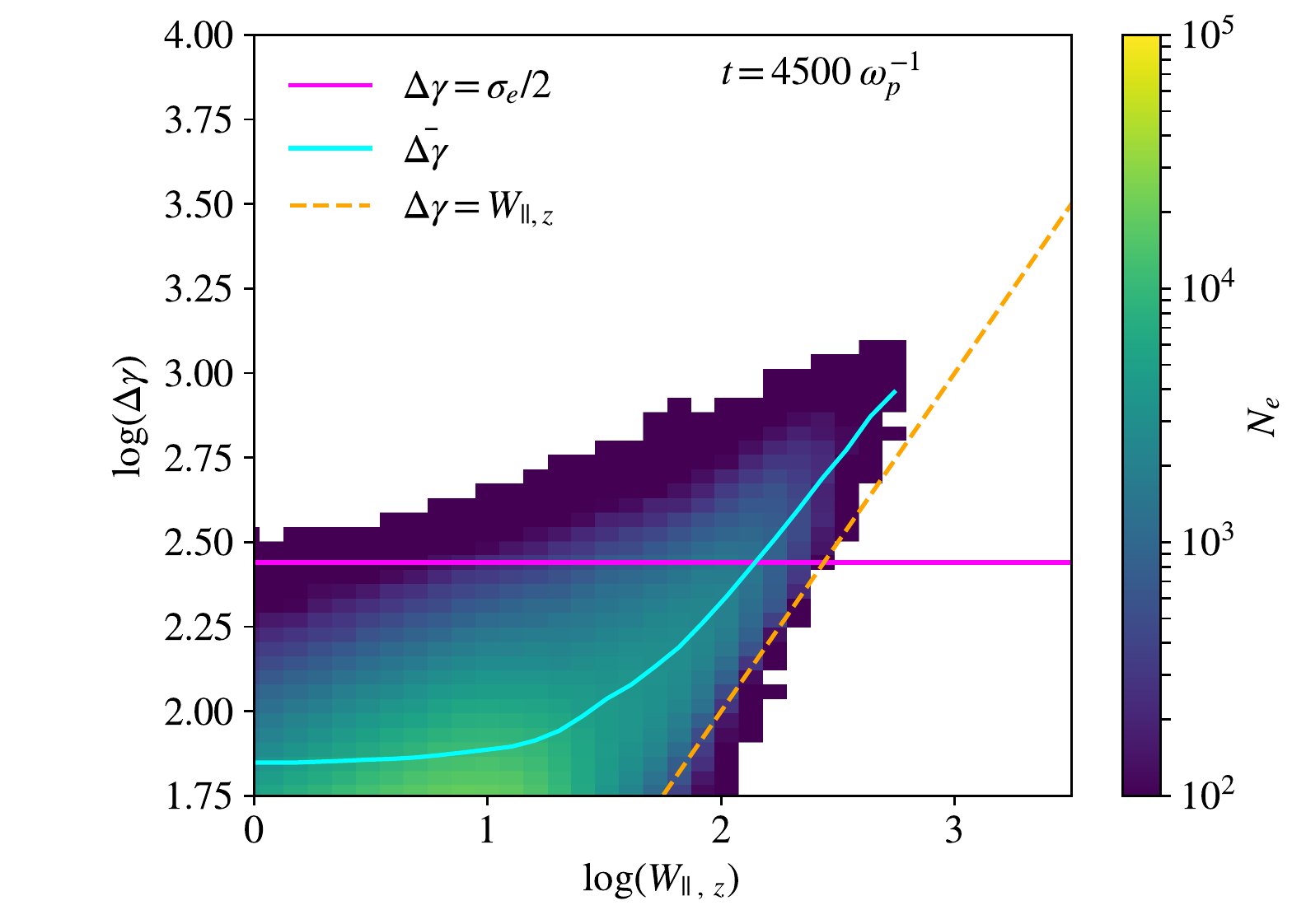}
	\newline
	\includegraphics[width=\linewidth]{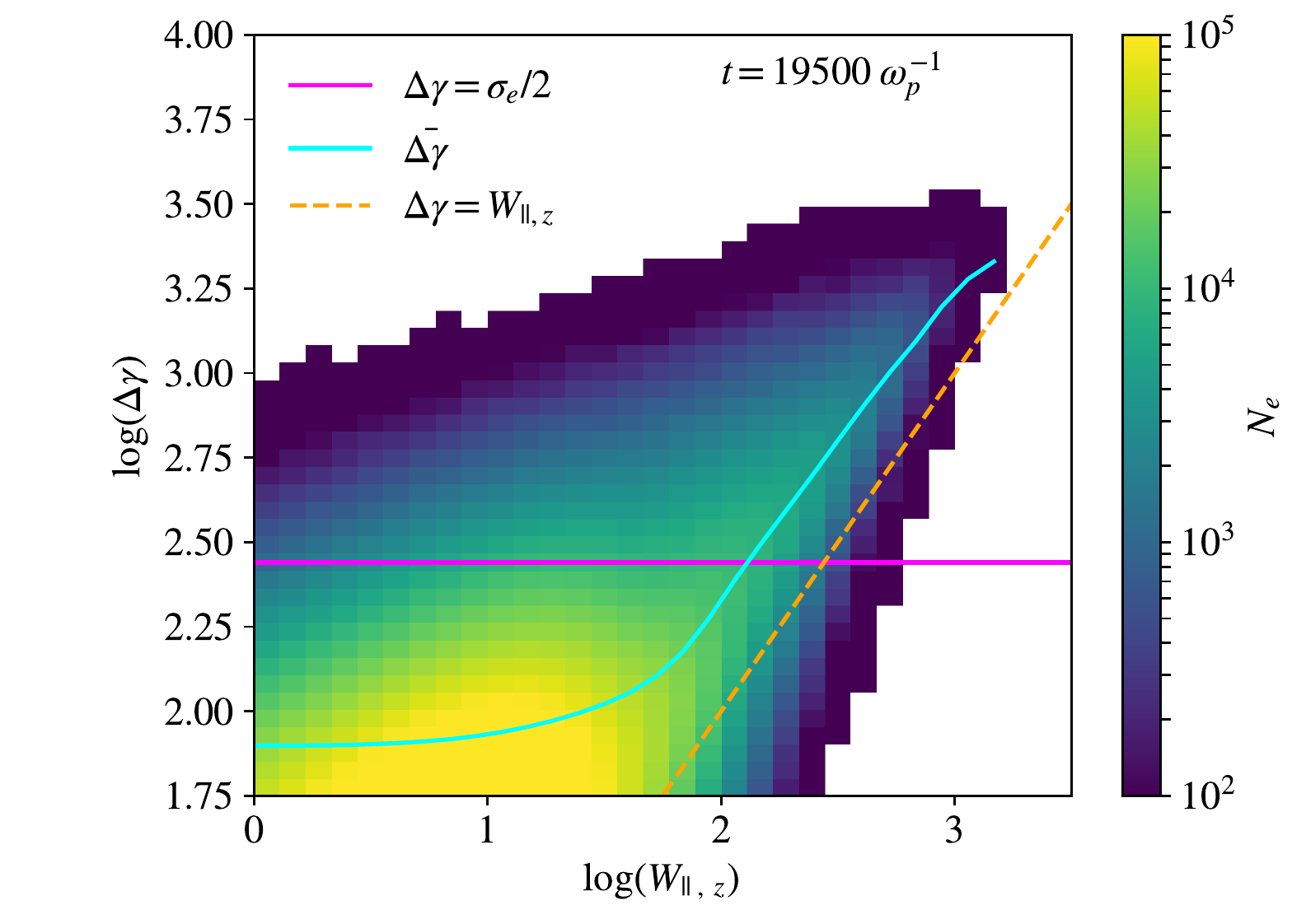}
	
	\caption{2d histograms of the $z$-component of parallel electric field work versus the total change in Lorentz factor taken at two different times (as indicated in the plot) from the triggered $B_{g}=0.1B_{0}$ simulation.  The cyan, magenta, and dashed orange lines have the same meaning as in \ref{wpar_hist_earlytime}.}
	\label{bguide1_wpar_hist}
\end{figure}

We show in Figure \ref{bguide1_wpar_hist} the same 2d histograms for the triggered simulation with $B_{g}=0.1 B_{0}$.  We see that there is still a strong correlation between electrons' final energy and the work done by the $z$-component of parallel electric fields: the non-ideal electric field at X-points is still playing an important role in accelerating a large number of non-thermal electrons.  However, we see that there is significantly more dispersion away from the $\Delta \gamma = W_{||,z}$ line than in Figure \ref{wpar_hist_earlytime}, where the secondary tearing mode is suppressed.  Most likely, this is because in an environment with an abundance of secondary plasmoid mergers, there are more available channels for energy gain that are not necessarily associated with $E_{||,z}$ (see, e.g., \citealt{guo2019}).  The untriggered counterparts to Figures \ref{wpar_hist_earlytime} and \ref{bguide1_wpar_hist} are nearly identical to the triggered cases, with a tight correlation between $\Delta \gamma$ and $W_{||,z}$ for the $B_{g}=0.3B_{0}$ case and a significant correlation but with more dispersion in the $B_{g}=0.1B_{0}$ case.
\\
\\

\section{Test Particles and The effect of parallel electric fields}\label{test_prtls}
In order to more thoroughly investigate our finding that non-thermal electron injection is largely controlled by  $E_{||,z}$ near X-points, we set up a test where we include populations of test electrons that only feel certain components of $\bold{E}_{||}$.  These electrons are evolved simultaneously with the normal particles but do not deposit their currents onto the grid.  We use two sets of test electrons to explore the energization mechanism of electrons.  One set of test electrons does not feel electric fields that are parallel to magnetic fields (i.e., $E_{||}=0$), and the other set does not feel any $z$-component (out-of-plane) of the $\boldsymbol{E_{||}}$ field ($E_{||,z}=0$).  

\begin{figure}[htp] 
\includegraphics[width=\linewidth]{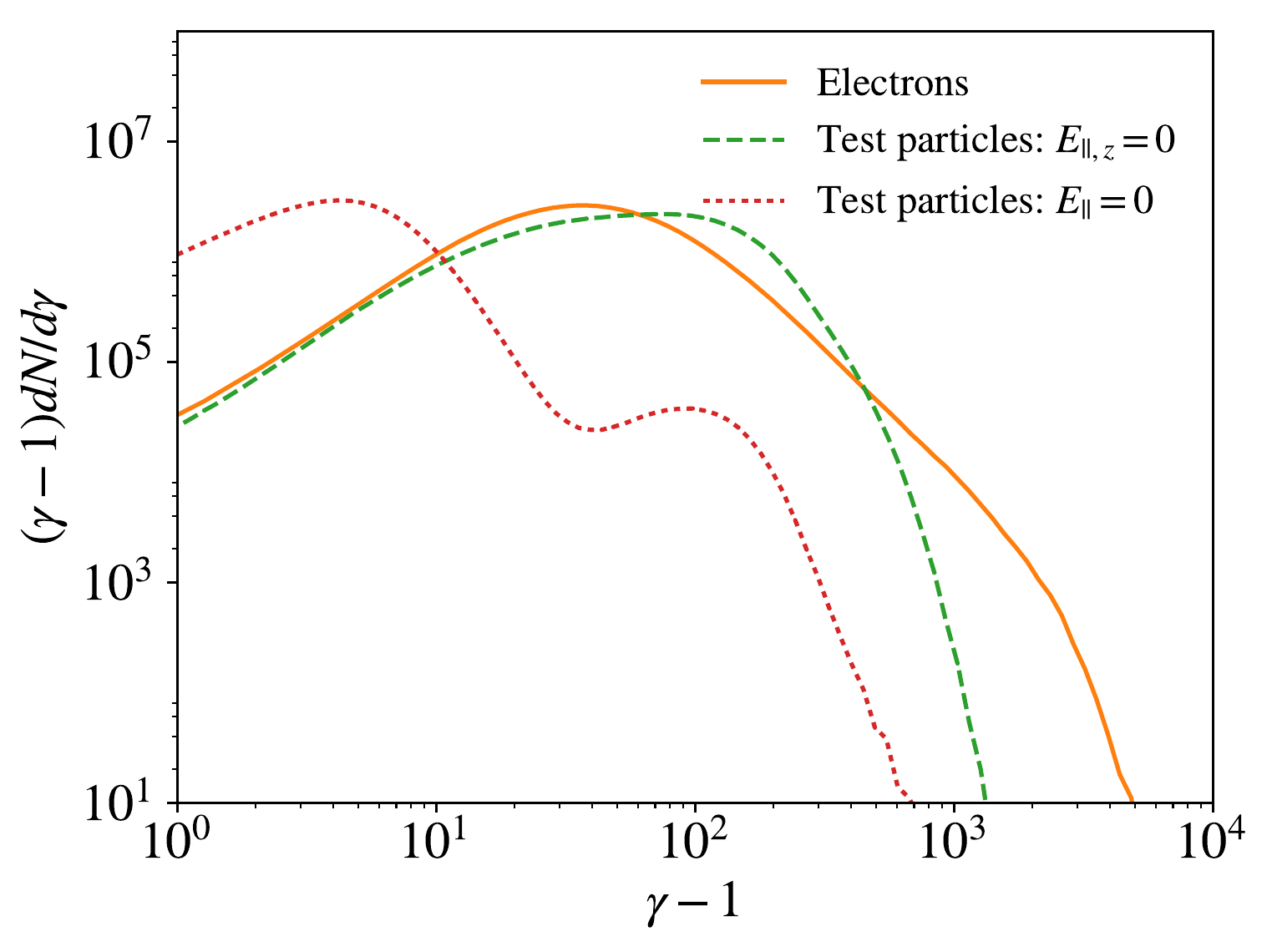}
\caption{Spectra of different populations of electrons in the triggered $B_{g}=0.3B_{0}$ simulation taken at $t=19500 \; \omega_{p}^{-1}$ taken only in the reconnection region.  The regular electron spectrum is shown in orange.  The green dashed line shows the spectrum of test electrons that do not experience parallel electric fields in the $z$-direction and the dotted red line shows the spectrum of electrons that do not experience any parallel electric fields.  All of the spectra are from the reconnection region (i.e., excluding the upstream plasma).  We see that the majority of test  parallel electric fields (red dotted line) are neither heated nor accelerated to non-thermal energies.  The test electrons that feel the in-plane component of parallel electric fields are heated to roughly the same overall thermal energy as the regular electrons but lack a power-law tail, indicating that the $z$-component of parallel electric fields is responsible for producing a non-thermal power law.}
\label{testprtl_spec}
\end{figure}

We show the spectra of these different populations of electrons in Figure \ref{testprtl_spec} for the triggered simulation with $B_{g}=0.3B_{g}$ at the last time in the simulation, $t=19,500 \; \omega_{p}^{-1}$ (i.e., given our periodic boundaries in $x$, the spectrum includes all the electrons processed by reconnection).  In this plot, the solid orange line shows the spectrum the regular electrons, the dashed green line shows the test electrons that do not feel the $z$-component of $\bold{E_{||}}$, and the dotted red line shows the test electrons that do not feel $\bold{E_{||}}$ at all.  All of the spectra shown are calculated only in the reconnection region.
We see the majority of electrons that do not experience any $\bold{E_{||}}$ (red dotted line) remain in their cold initial distribution with $\gamma \approx 4$.  A small number of these electrons (about $1/1000$ of the electrons in the reconnection region) are heated by $\boldsymbol{E_{\bot}}$ to typical thermal energies of the regular electrons of $\gamma \approx 100$.  This demonstrates that both bulk heating and non-thermal acceleration are mediated by parallel electric fields.  The electrons that feel the in-plane parallel fields $E_{||,xy}$ but not the $z$-component (green line), however, have a thermal peak that is roughly consistent with the normal electrons at $\gamma \approx 100$, but shows no evidence of a non-thermal distribution above this thermal peak in their spectrum.  This confirms our conclusion that the non-thermal tail of the distribution is controlled by the parallel electric field in the $z$-direction associated with X-points.  The thermal peak is still able to form because electrons that enter the current layer in the outflow (i.e., far from the primary X-point) primarily interact with the outflow-aligned electric field, $E_{||,x}$, which heats them up to $\gamma \sim 100$.

\begin{figure}[htp] 
	\includegraphics[width=\linewidth]{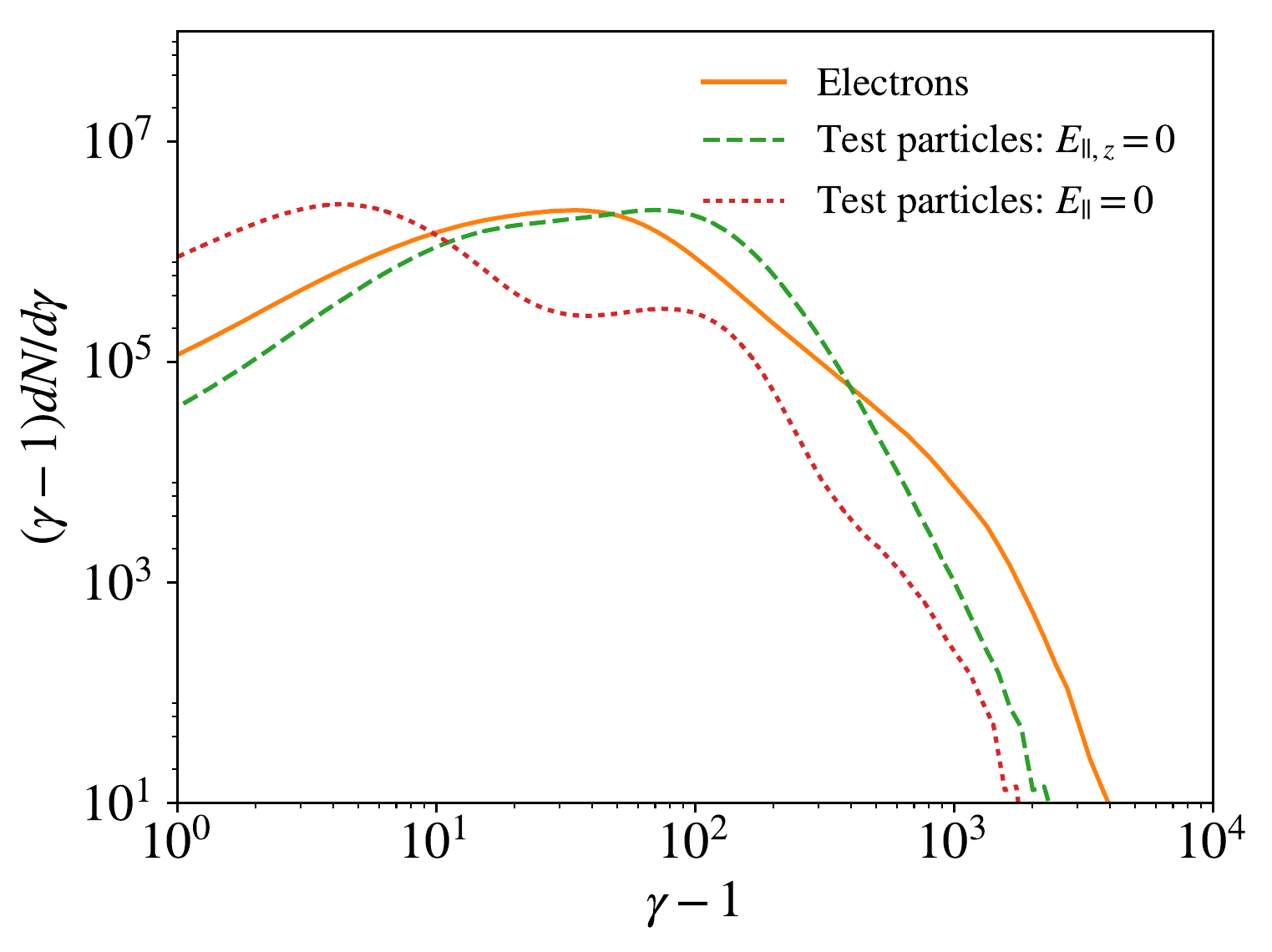}
	\caption{Spectra of various populations of test electrons from the triggered simulation with $B_{g}=0.1B_{0}$ taken at $t=19500 \; \omega_{p}^{-1}$ taken only in th reconnection region.  Again we see that the electrons that do not feel the $z$-component of parallel electric fields (dashed green line) are heated to similar temperatures as the regular electrons, but lack the hard non-thermal tail above the thermal peak.  }

	\label{bguide1_testprtl_spec}
\end{figure}

In Figure \ref{bguide1_testprtl_spec}, we show the spectra of test electrons for the triggered $B_{g} = 0.1 B_{0}$ simulation.  We see that compared to the $B_{g}=0.3B_{0}$ case, a larger number of electrons that don't feel any parallel electric fields are heated to the thermal bump at $\gamma \approx 100$.  We argue that this is because in an environment where the secondary tearing mode is active, plasmoids continually form and merge, providing more sites where plasma motions, plasmoid contraction, and other processes can heat up the electrons via $\boldsymbol{E_{\bot}}$.  

We also see with the second group of test particles, i.e., the electrons that feel $E_{||,xy}$ but not the $z$-component again reach energies comparable to the thermal peak of the spectrum of the normal electrons, indicating that the x and y components of $\bold{E}_{||}$ are important to the overall heating process.  In this population of electrons, the power-law slope (if any) is significantly softer than for the regular electrons.  We again clearly see that if the $z$-component of the parallel non-ideal electric field were to be absent, then the strong signature of non-thermal electron acceleration would not appear.

These experiments with test particles demonstrate the importance of non-ideal fields at X-points in the injection of high energy particles.  In terms of the overall energy budget (i.e., electron heating), the non-ideal electric field associated with X-points, $E_{||,z}$, is clearly not the dominant driver: the electrons that do not feel $E_{||,z}$ (dashed green line) have a comparable average Lorentz factor to the regular electrons (orange line).  However, the populations of electrons that do not feel $E_{||,z}$ have a significantly softer spectrum with little to no evidence of any non-thermal component.
	
\section{conclusions}
In this work, we have investigated electron acceleration mechanisms in trans-relativistic reconnection with a set of four 2.5 dimensional PIC simulations.  We include acceleration diagnostics that are calculated on-the-fly for all of the particles so that our results are not affected by time or particle downsampling of output files, or biases introduced by only looking at a small number of particle trajectories.

We dissect the most important ingredients for acceleration by varying the number of plasmoids and X-points.  We attain this by changing the simulation setup or the guide field strength.  In particular, we choose either a triggered setup (where we induce reconnection by hand at a single location in the current sheet) or an untriggered setup (where reconnection evolves spontaneously).  In the triggered setups a primary X-point forms and results in outflows away from the middle of the box such that the only plasmoid mergers that occur are secondary plasmoids that are generated via the secondary tearing mode.  In untriggered setups, the current sheet pinches at multiple places, resulting in a chain of primary X-points and plasmoids.  These primary plasmoids will inevitably merge in a hierarchical manner, resulting in numerous large plasmoid mergers.  

The physical parameter we vary is the strength of the guide field.  For our values of $\sigma=0.3$ and $\beta_{i}=0.003$, a modest guide field of $B_{g}=0.3B_{0}$ suppresses the secondary tearing mode resulting in a reconnection layer dominated by outflows and primary X-points.  For our lower value of $B_{g}=0.1B_{0}$, however, the secondary tearing mode is active, which fragments the layer into a series of secondary X-points and plasmoids.

We show that X-points in both the primary current sheet and merger-induced current sheets are the dominant sites of electron injection due to the strong $E_{||,z}$, the (non-ideal) z-component of the parallel electric field, that occurs at these locations.  We show this by classifying the location of the first acceleration episodes of all of our electrons.  We find that the electron energy spectra above $\sigma_{e}/2$ are dominated by electrons that are injected near an X-point for all of our simulations.  In the cases where the secondary tearing mode is suppressed ($B_{g}=0.3B_{0}$), X-points in the primary current sheet are responsible for injecting the vast majority of electrons that ultimately exceed $\gamma > \sigma_{e}/2$.  In contrast, when the secondary tearing mode is active there are copious plasmoid mergers and X-points in the MCS dominate the injection of high energy electrons.

Furthermore, we show that the work done by parallel electric fields in the $z$-direction, $W_{||,z}$, play a critical role in relativistic electron acceleration.  By tracking the work done by the $z$-component of parallel electric fields on-the-fly for all of our particles, we show that for electrons that end with $\gamma > \sigma_{e}/2$, there is a strong correlation between the final Lorentz factor and $W_{||,z}$.  The correlation is especially tight in the $B_{g}=0.3B_{0}$ case, but is still significant in the lower guide field case, where there are more acceleration channels due to the presence of numerous secondary plasmoids.

We further demonstrate the importance of $E_{||,z}$ by tracking populations of test electrons that are evolved alongside the regular particles in the simulations but do not deposit their currents onto the grid.  We prescribe that the test electrons only feel certain components of $\bold{E}_{||}$.  Using these test electrons, we confirm that $E_{||,z}$ is primarily responsible for regulating the high-energy non-thermal tail of the distribution, while the other components of $\boldsymbol{E_{||}}$ (mostly $E_{||,x}$ along the boundaries of the outflow) are largely associated with bulk heating.  

Numerous studies have explored particle acceleration mechanisms in magnetic reconnection (e.g., \citealt{zenitani2001, nalewajko2015, guo2015, guo2019, dahlin2015}).
The importance of the X-point phase in injecting electrons into a non-thermal distribution has been highlighted by \citet{zenitani2001} and  \citet{nalewajko2015}.  \citet{dahlin2015} also found in the non-relativistic case, that Fermi processes associated with the curvature drift play a dominant role when the guide field is low and that parallel electric fields become more important as the guide field increases.  This study, however, was focused on electron bulk heating rather than the acceleration of the relatively few non-thermal electrons, for which the dominant energization mechanism may be completely different than for the majority of electrons, as we indeed find.

We expand upon these previous works in a number of ways.  First, we employ the true electron-proton mass ratio in all of our simulations. Second, we vary the number of X-points by the choice of whether or not to trigger reconnection, as well as box size tests and sheet thickness tests that support our conclusions (see Appendix \ref{box_size} and \ref{thickness}).  Third, we test directly the effect of parallel electric fields by tracking populations of test particles that do not feel certain components of $\boldsymbol{E_{||}}$.

We note that our conclusions are significantly different than recent work by \citet{guo2019}, who argued for a negligible role of non-ideal fields.  
We can understand these differences in the context of how the energy gains associated with X-point and Fermi acceleration scale with $\sigma_{e}$ and $\sigma$.  The electron energy gain ($\Delta \gamma$) at an X-point is roughly proportional to $\sigma_{e}$, while the Alfv\'en speed, and hence speed of the scattering centers in the Fermi process goes as $v_{A}/c=\sqrt{\sigma/(\sigma+1)}$.  If an electron is entrained in the outflow, it will be at the very least energized to the outflow's bulk Lorentz factor, $\Gamma=\sqrt{\sigma+1}$.  Alternatively, the first time an electron scatters off the outflow, it may be energized up to $\sim \Gamma^{2}=\sigma+1$, analogous to a particle's first scattering event off of a relativistic shock (\citealt{achterberg2001})\footnote{For this argument, we have assumed $\Gamma\gg1$, for the sake of simplicity.}.  

In our study, we use $\sigma=0.3$, which makes the outflow only mildly relativistic ($v_{A}/c=0.48$).  Additionally, we employ the true electron-proton mass ratio, such that $\sigma_{e} \simeq 550$.  As such, the typical change in Lorentz factor at an X-point in our setup is much greater than the energy of an electron that scatters off the outflow ($\Gamma^{2} = 1.3$).   In contrast, \citet{guo2019} employs $\sigma=50$ (i.e., $v_{A}/c=0.99$) and a pair plasma such that $\sigma_{e}=100$.  In this regime, the change in Lorentz factor at an X-point is no longer as dominant over Fermi-type processes ($\Gamma^{2}=51$).  When Fermi-type processes are able to compete with energization at X-points, then Fermi-type acceleration can naturally dominate both the injection and acceleration of electrons.  This is because the volume of the outflow is much larger than the volume occupied by X-points.


Our findings emphasize the importance of understanding the structure and dynamics of current sheets when applying a prescription for electron acceleration in models of astrophysical systems.  In particular, if a significant guide field is present, then the secondary tearing mode is suppressed and the efficiency of electron acceleration may decrease dramatically.  If such a current sheet is thick, then the primary tearing mode generates a small number of X-points and electron acceleration will be negligible.  In such a case, one must carefully understand the dynamics of the flow around the current sheet: if external perturbations are present, then the current sheet may still tear and develop numerous primary X-points (as in our untriggered simulations), resulting in significant acceleration, but if not, then a thick current sheet with a modest guide field will have almost no high-energy electron acceleration occur during reconnection.
	
Electron acceleration in reconnection is invoked to explain the variability and hard spectra of numerous astrophysical systems, including low-luminosity accretion flows, blazars, and pulsar wind nebulae.  We find that the conditions necessary for efficient electron injection and acceleration may be present in some systems, but cannot be universally assumed.  In addition to having to consider the physical properties (magnetization, plasma-beta, guide field strength, width) of current sheets, our results point to the importance of understanding the formation and dynamics of current sheets in a macroscopic context (e.g., \citealt{uzdensky2016}).  We suggest a careful timescale analysis when invoking reconnection in a macroscopic system, considering the formation of the current sheet, the growth time of the primary tearing mode, and the timescale of dynamical disruptions that may inhibit (or potentially trigger and enhance, depending on the nature of the disruption) reconnection.

Our results further the understanding of non-thermal electron acceleration in the trans-relativistic regime of reconnection, which is potentially important for explaining the observed properties of nearby radiatively inefficient accretion flows as well as the X-ray flares observed from Sgr~A*.   Pinning down the electron acceleration mechanisms in this regime can also inform efforts to include non-thermal electrons in models of accretion flows.  Ultimately, this understanding may lead to a physically grounded model for electron acceleration via reconnection in astrophysical sources.

\section*{Acknowledgments}

We gratefully acknowledge support for this work from NSF AST-1715061 and Chandra Award No. TM6-17006X.   LS acknowledges support from DoE DE-SC0016542, NASA ATP NNX-17AG21G, and
NSF ACI1657507.  The simulations were performed on El Gato at the University of Arizona and NASA High-End Computing (HEC) program through the NASA Advanced Supercomputing (NAS) Division at AMES Research Center.

\appendix

\section{Appendix A: Dependence on Box Size}\label{box_size}
In this appendix we explore how the electron spectra of our simulations depend on the length of the simulation domain along the current sheet (in the $x$-direction).  As previously discussed, the hardness of the non-thermal tail depends on the number of X-points per unit length: if there are more X-points in a current sheet of a given length, then we expect the slope to be harder.  As a result, we expect the spectra from triggered simulations with $B_{g}=0.3B_{0}$ to be the most dependent on the size of the box.  Because there is only one primary X-point, the number of X-points per unit length scales as we vary the box size as $1/L_{x}$, where $L_{x}$ is the length of the current sheet.  

\begin{figure}[htp] 
	\includegraphics[width=\linewidth]{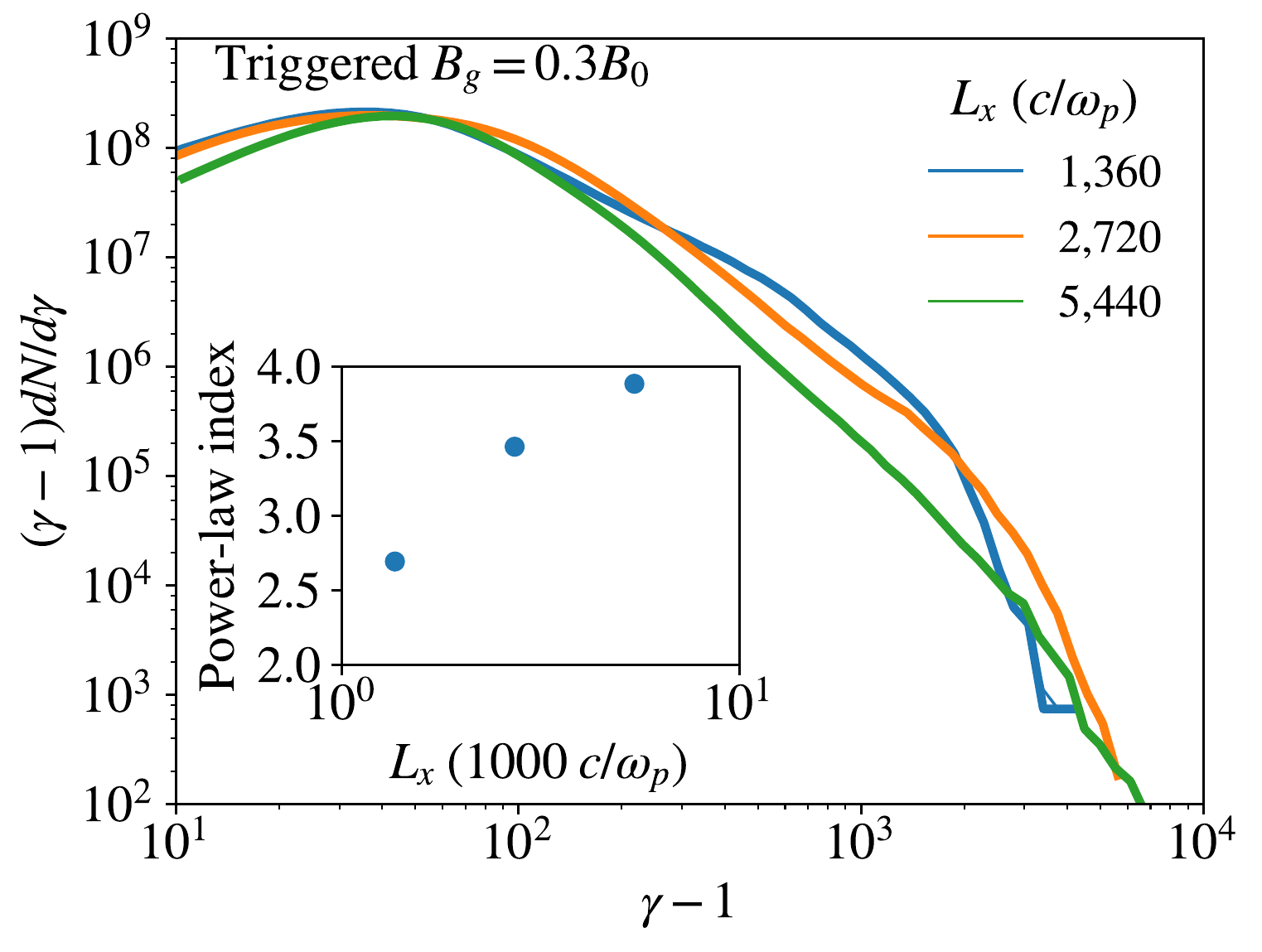}
	\caption{Spectra from triggered simulations with $B_{g}=0.3B_{0}$ and with varying box length along the current sheet (runs A0*, A1, A2).  Spectra are normalized such that their thermal peaks have a comparable number of electrons.  We see that the spectra steadily become softer as the box length increases.  The inset shows the power-law index vs. box length.}
	\label{triggered_bg_3_boxsize}
\end{figure}

We show in Figure \ref{triggered_bg_3_boxsize} the spectra from simulations with this setup of varying box sizes.  In the inset we plot the length of the box (along the current sheet) vs the measured power-law index.  As expected, we see that the spectra steepen significantly as the length of the current sheet increases.  This is because the single primary X-point mediates the vast majority of high-energy electron injection, so for a larger box, a particle is less likely to interact with the X-point, and more likely to interact with the outflow, resulting in a smaller fraction of electrons in the non-thermal tail as compared to the thermal peak.

\begin{figure}[htp] 
	\includegraphics[width=\linewidth]{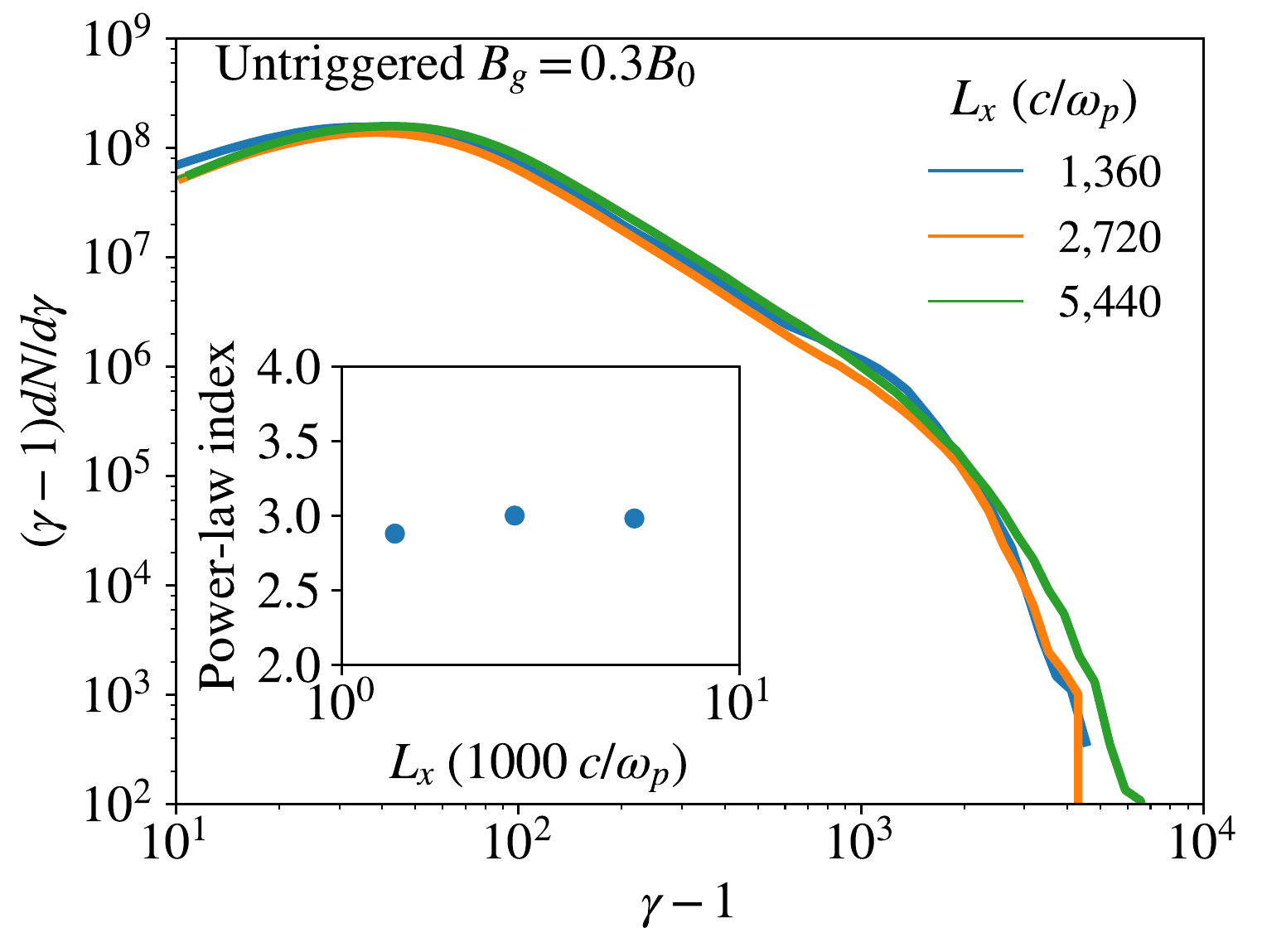}
	\caption{Spectra from untriggered simulations with $B_{g}=0.3B_{0}$ and with varying initial box length along the current sheet (runs B0*, B1, B2).  In this case, the shape of the spectrum is nearly independent of box size.}
	\label{untriggered_bg_3_boxsize}
\end{figure}

In an untriggered setup, however, the number of primary X-points per length of the current sheet is fixed by the dominant mode of the primary tearing instability, which depends on the current sheet thickness but not on the box length.  As such, the probability that a given electron interacts with an X-point versus somewhere along the rest of layer should be roughly independent of the box size, and hence we expect the spectra to correspondingly be insensitive to the box size for an untriggered setup.  We show in Figure \ref{untriggered_bg_3_boxsize} the spectra from untriggered simulations with $B_{g}=0.3B_{0}$.  We see that the power-law index of the high-energy tail is almost independent of the length of the box.

\begin{figure}[htp] 
\includegraphics[width=\linewidth]{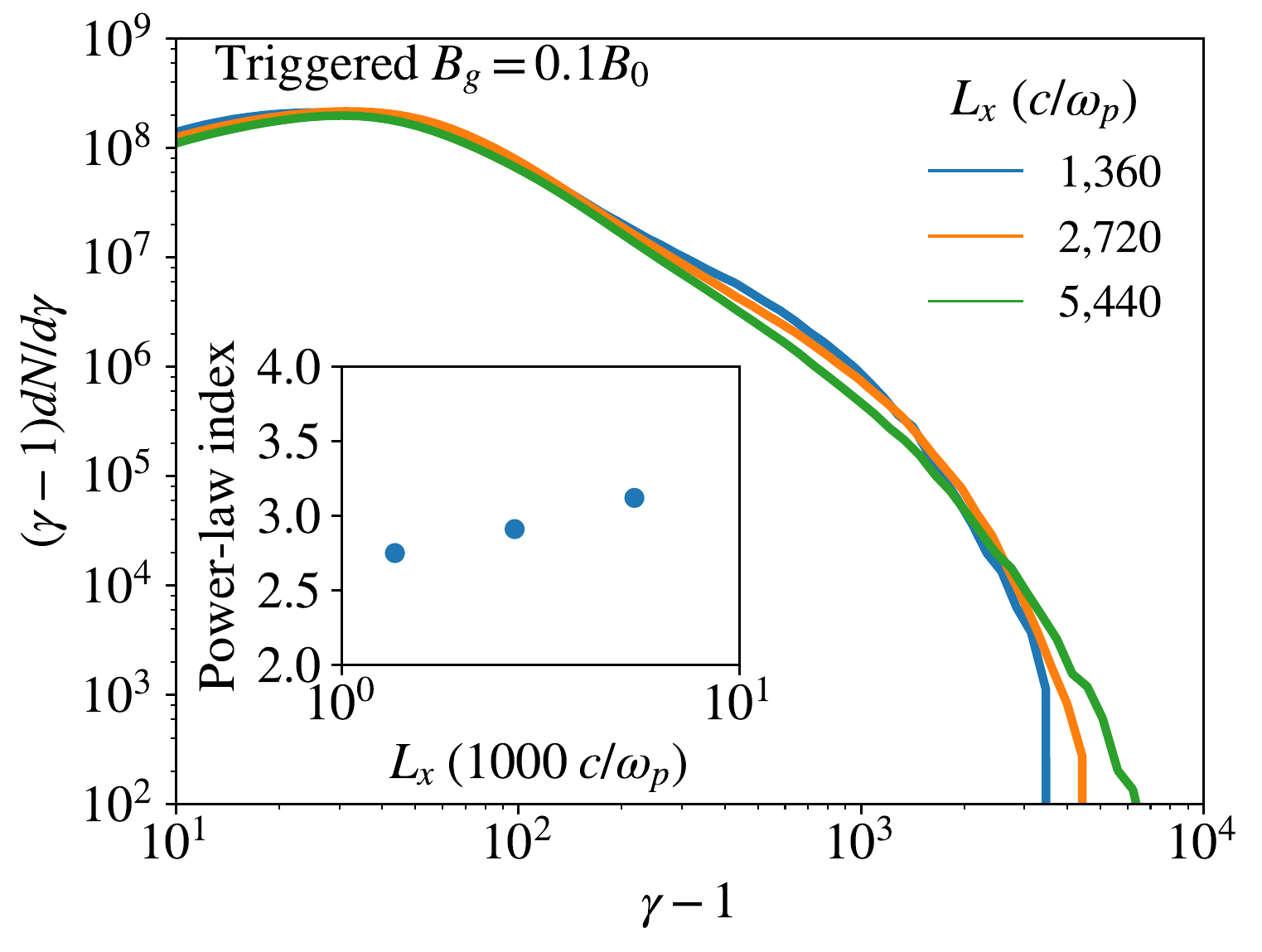}
\caption{Spectra from triggered simulations with $B_{g}=0.1B_{0}$ and with varying box lengths (runs C0*, C1, C2).  We see that the power-law slopes depend on the initial box length, but not nearly as strongly as in the triggered $B_{g}=0.3B_{0}$ case.}
\label{triggered_bg_1_boxsize}
\end{figure}

We show in Figure \ref{triggered_bg_1_boxsize} the spectra from triggered simulations with a guide field strength of $B_{g}=0.1B_{0}$.  We find that the electron spectra depend on the initial box length, but much more weakly than in the triggered case with $B_{g}=0.3B_{0}$ shown in Figure \ref{triggered_bg_3_boxsize}.  This is expected because there are numerous secondary X-points along the current layer, so the number of X-points per unit length of the sheet does not simply scale as  $1/L_{x}$.  Furthermore, in the limit of a very large domain, we do not expect acceleration to be negligible: when the secondary tearing mode is active, X-points will invariably be present along the entire current layer and will not represent a vanishingly small fraction of the total length of the current sheet, as it happens in the triggered case where the secondary tearing mode is suppressed.

\begin{figure}[htp] 
	\includegraphics[width=\linewidth]{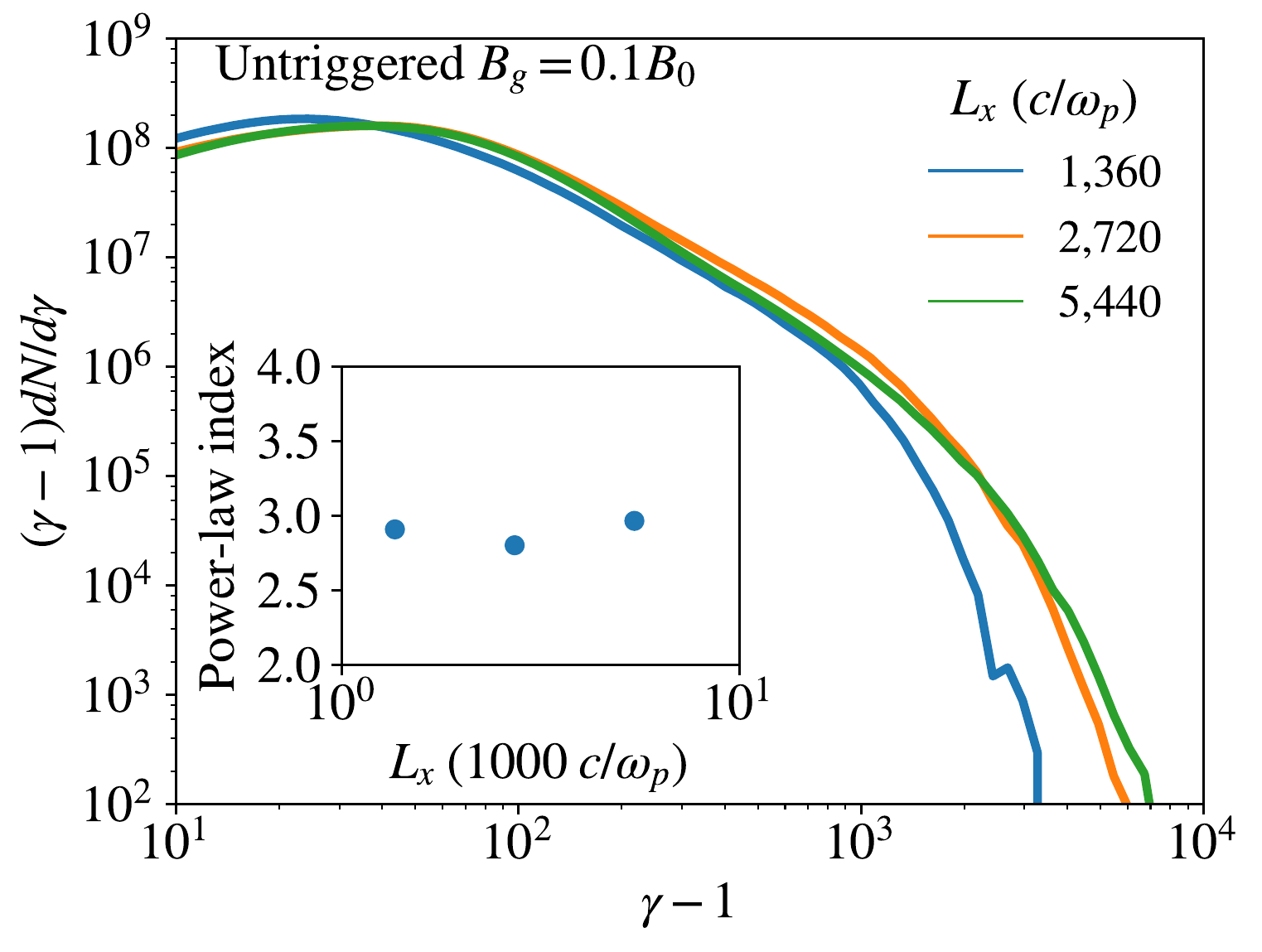}
	\caption{Spectra from untriggered simulations with $B_{g}=0.1B_{0}$ and with varying box lengths (runs D0*, D1, D2).  The power-law slopes do not have a systematic dependence on box size.}
	\label{untriggered_bg_1_boxsize}
\end{figure}
We show in Figure \ref{untriggered_bg_1_boxsize} the spectra from untriggered simulations with $B_{g}=0.1$ at varying box sizes.  We see, as expected, that the high-energy slope has no discernible dependence on box size.

\section{Appendix B: Effects of Initial Sheet Thickness}\label{thickness}
\begin{figure*}[htp] 
	\includegraphics[width=\linewidth]{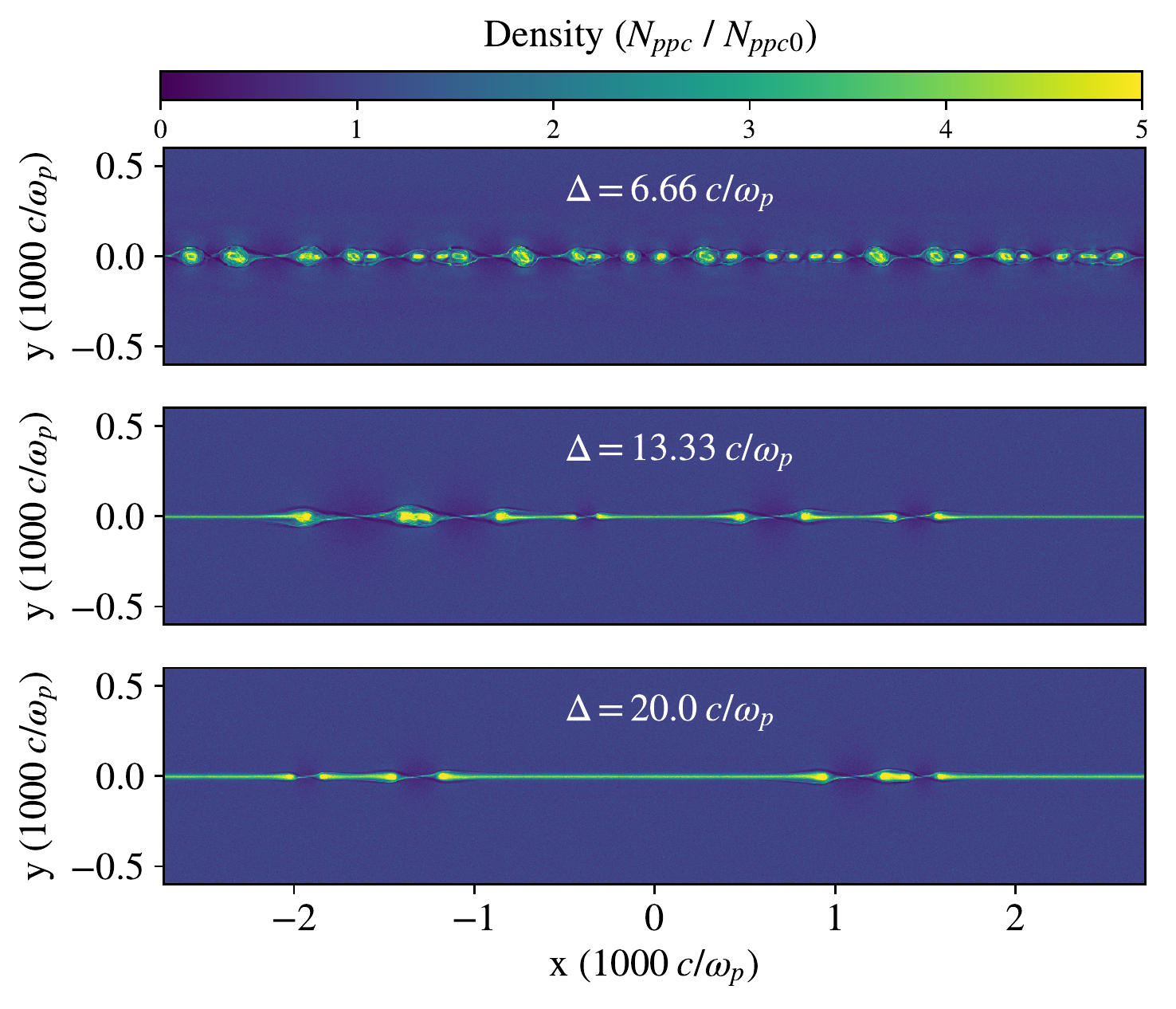}
	\caption{Density structures of three untriggered simulations (runs B2, B3, B4) with $B_{g}=0.3{B_{0}}$ with varying initial current sheet widths.  These snapshots are taken at different times in each simulation, corresponding to the time when the primary tearing mode has just finished developing the primary X-points.  These time are at $t=1200, \; 2400, \; 15000 \; \omega_{p}^{-1}$ for the $\Delta=6.66, \;13.33, \;20 \; c/\omega_{p}$ cases, respectively.}
	\label{untriggered_thickness_flds}
\end{figure*}

\begin{figure}[htp] 
	\includegraphics[width=\linewidth]{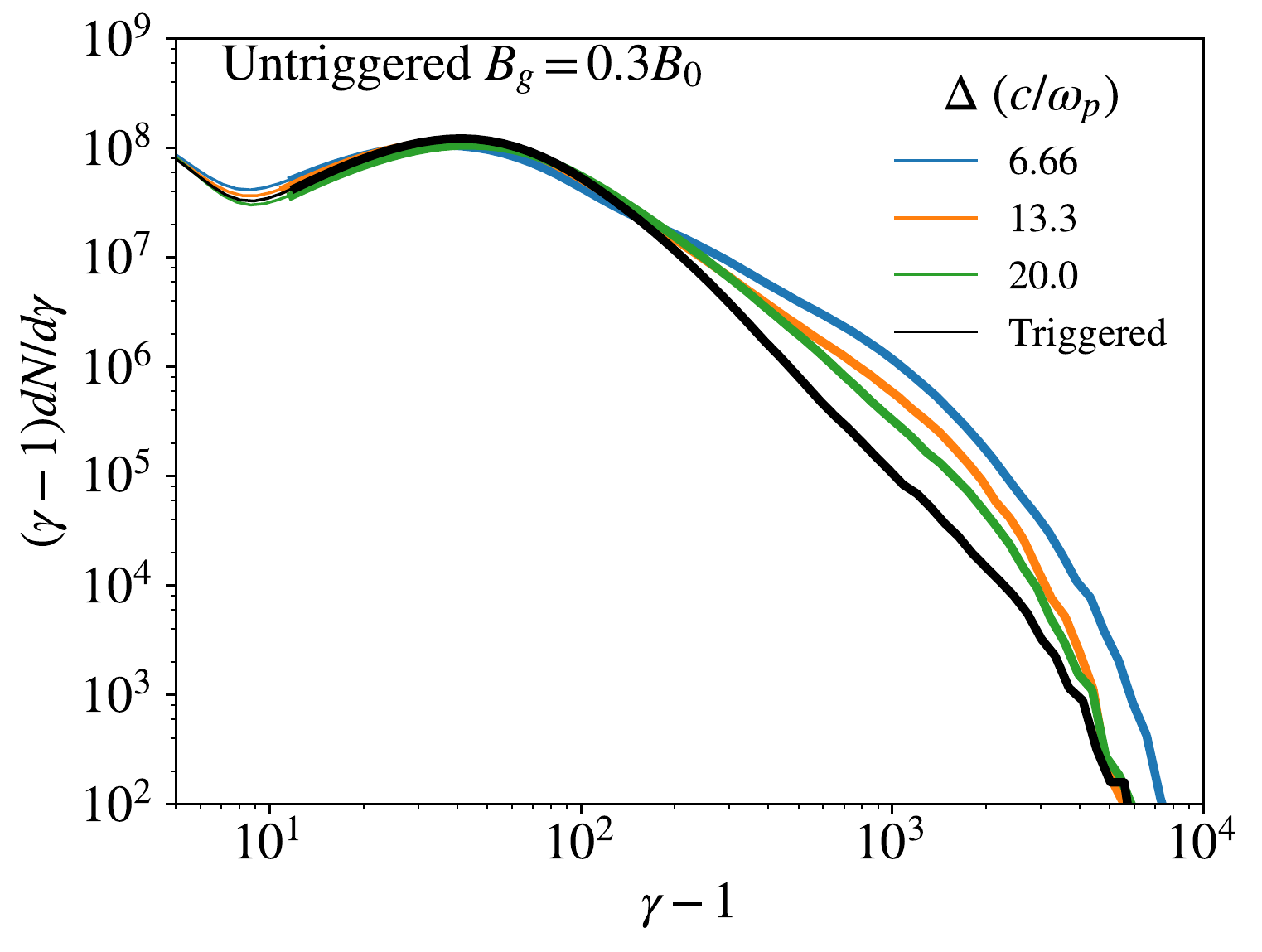}
	\caption{Electron energy spectra from three $B_{g}=0.3B_{0}$ untriggered simulations with different initial sheet thicknesses as well as the triggered simulation (runs B2, B3, B4, and A2).  As the initial width of the current sheet increases, fewer X-points spontaneously form via the primary tearing mode, and the spectra soften accordingly,  eventually approaching the limit of a triggered simulation (which corresponds to the case of a single primary X-point)}
	\label{untriggered_thickness_spec}
\end{figure}

The initial thickness of the current sheet, $\Delta$,  in an untriggered setup controls the number of primary X-points that develop in the current sheet.  For currents sheets of a fixed length, $L_{x}$, we expect relatively few primary X-points to develop in a thick sheet, whereas thin sheets fragment copiously into a chain of secondary plasmoids.  As we have shown in this paper, the number of X-points per unit length affects the resulting non-thermal electron spectrum: current sheets with more X-points and plasmoids tend to have harder spectra.  We further test this assertion here by varying the initial thickness of the current sheet in our untriggered $B_{g}=0.3B_{0}$ setup.  We show in Figure \ref{untriggered_thickness_flds} snapshots of density for three simulations with different initial sheet thicknesses.  For this comparison, we use a simulation domain that is twice as long as in our fiducial setups to explore a wide range of X-points per unit length.  We show each snapshot at a time right after the primary X-points have developed.  As such, we show each simulation at a different physical time because the timescale of the primary tearing mode scales depends on $\Delta$ (see e.g., \citealt{brittnacher1995}).  As expected, we see that the thinnest sheet fractures in a multitude of X-points and plasmoids and the thicker sheets have correspondingly fewer primary X-points (five X-points in the middle panel for our fiducial sheet thickness, and four X-points for an initial width of $20 \; c/\omega_{p}$).  
We show in Figure \ref{untriggered_thickness_spec} the electron energy spectra from these three simulations.  As expected, we see that the thinnest sheet has the hardest spectrum, and the spectra soften as the initial width of the current sheet increases.

We note that in the limit of a very thick current sheet, there will be only one primary X-point, and the result should be identical to a triggered simulation with the same physical conditions.  The triggered simulations hence represent the thick-sheet limit of the untriggered setup and, as such, set the lower-limit on electron acceleration in a domain of a given size.  Indeed we see this is the case in Figure \ref{untriggered_thickness_spec}: the triggered simulation (black line) has a significantly softer spectrum than the untriggered counterparts with multiple primary X-points.

\section{Appendix C: Anti-parallel and Low Guide Field Comparison}
\begin{figure}[htp] 
	\includegraphics[width=\linewidth]{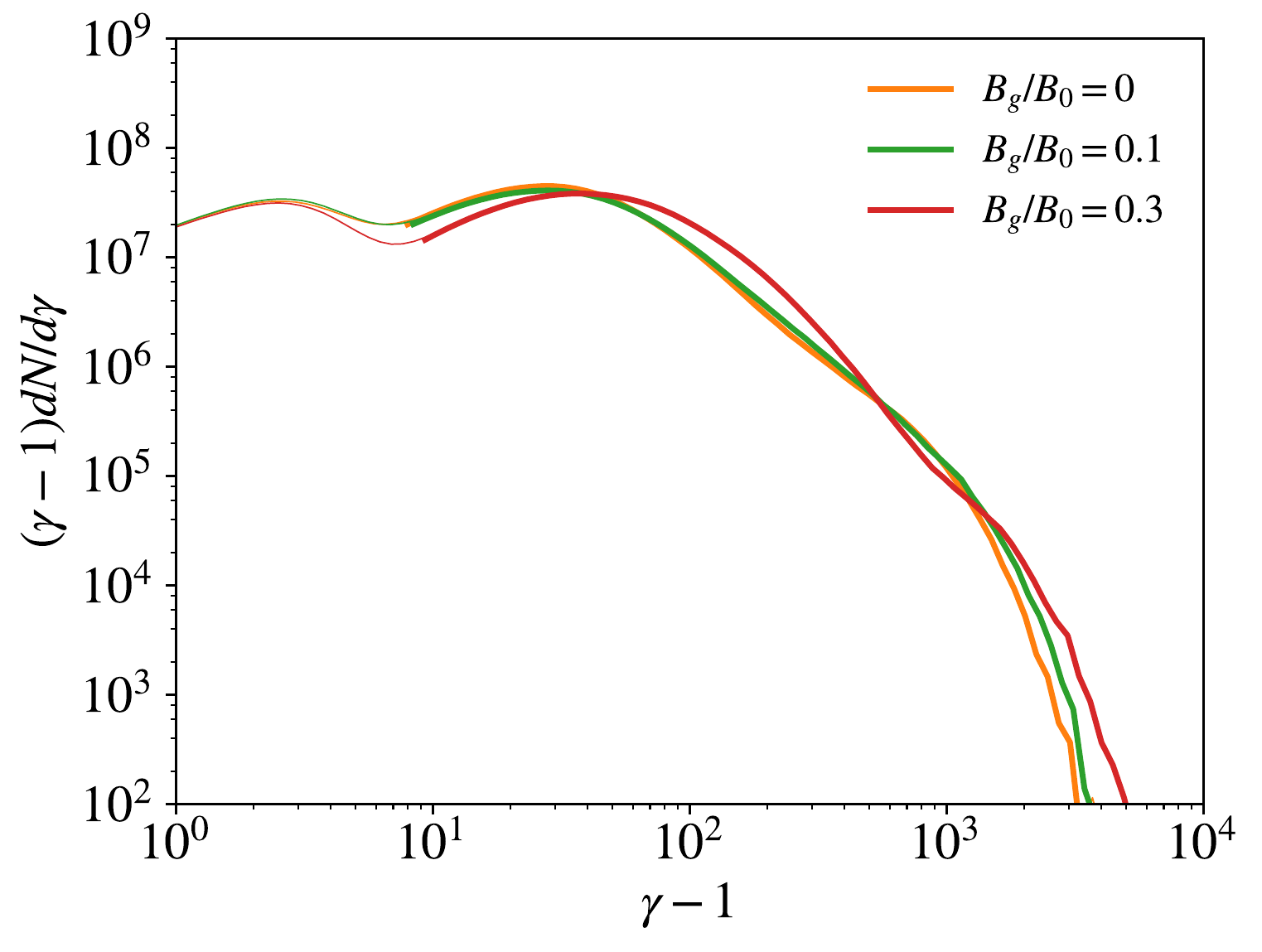}
	\caption{Electron energy spectra for triggered simulations with varying guide field and our fiducial box size (simulations A0*, C0*, and E0*).  The zero guide field case is shown in orange and $B_{g}=0.1B_{0}$ and $B_{g}=0.3B_{0}$ are shown with the green and red lines, respectively.  We see that the zero guide field and $B_{g}=0.1B_{0}$ spectra are nearly identical, while the $B_{g}=0.3B_{0}$ case shows more overall heating, but a steeper spectrum.}
	\label{bg_spec_compare}
\end{figure}

\begin{figure*}[htp] 
	\includegraphics[width=\linewidth]{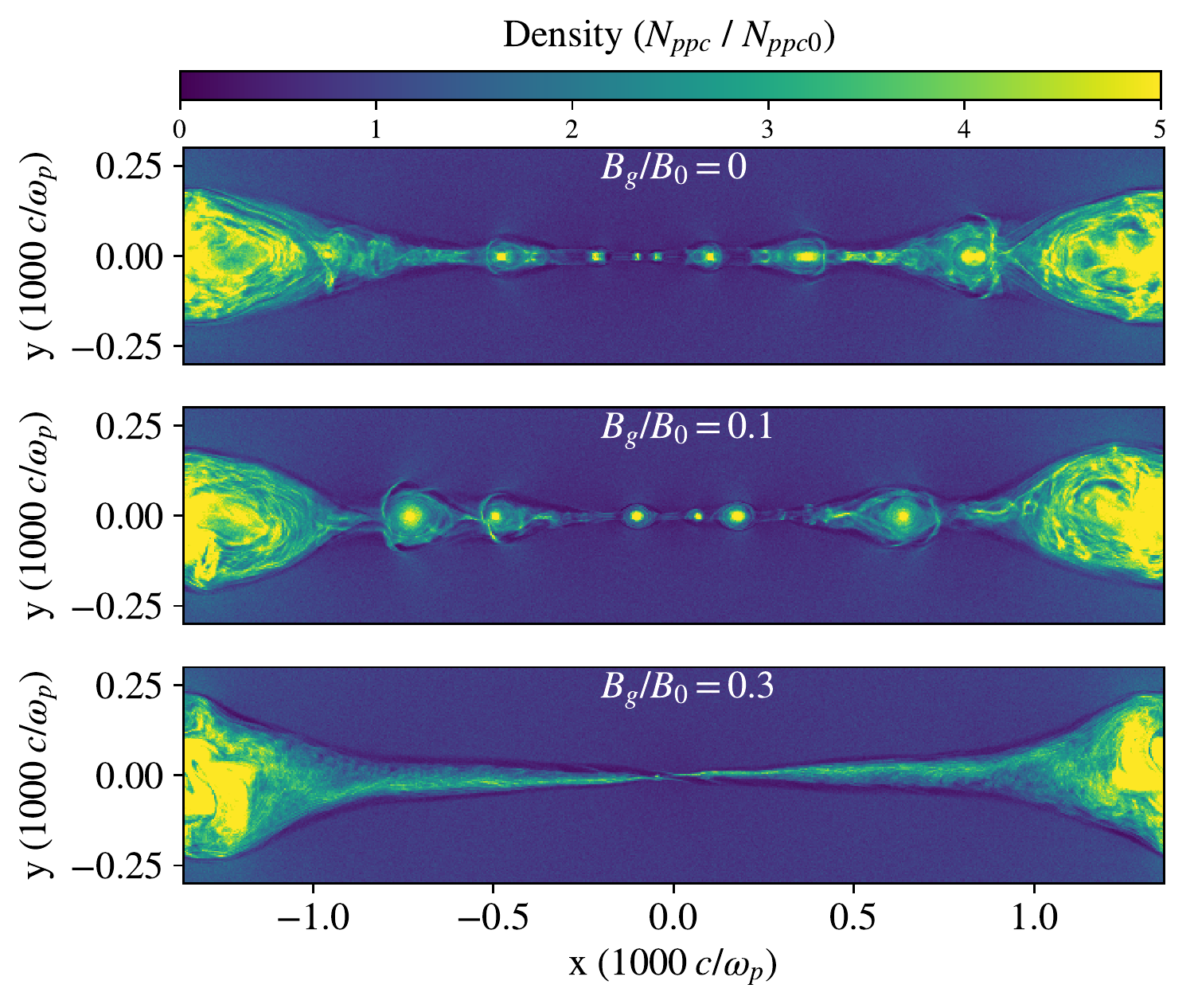}
	\caption{Snapshots of density for simulations with varying guide field.  We see that the structures in the purely anti-parallel case (top) and the weak guide field case ($B_{g}=0.1$) are remarkably similar; plasmoid and X-point formation occur copiously via the secondary tearing mode throughout the reconnection layer.  The higher guide field case (bottom) only shows a single X-point  tearing mode is suppressed.}
	\label{bg_flds_compare}
\end{figure*}

In this paper, all of the simulations we use have a non-zero guide field so that we can properly define $\boldsymbol{E_{||}}$ and compute $W_{||,z}$.  In the case of anti-parallel reconnection, this quantity is not well defined at X-points because the magnetic field is zero at these locations. So, even though non ideal fields are still present, they cannot be captured as $E_{||}$.  In this appendix, we show that our low-guide field ($B_{g}=0.1B_{0}$) results are similar to the zero guide field case.  We show in Figure \ref{bg_spec_compare} the spectra from simulations with identical physical parameters except for a changing guide field, with values of $B_{g}/B_{0}=0, \; 0.1\; $ and $0.3$.  We see that the spectra from $B_{g}/B_{0}=0$ and $B_{g}/B_{0}=0.1$ cases are remarkably similar, with nearly identical power-law slopes. Because of this, we argue that our conclusions for $B_{g}=0.1 B_{0}$ could also be applicable to the special case of anti-parallel reconnection ($B_{g}=0$).  We additionally examine the typical structures present in each simulation, shown in Figure \ref{bg_flds_compare}.  We see that in the two lowest guide field cases numerous X-points and plasmoids form, which ultimately result in similar spectra.  In the $B_{g}=0.3B_{0}$ case, however, the structure of the current layer changes dramatically, and these differences are reflected in the spectrum.  

\FloatBarrier

\bibliography{david_bib}
\bibliographystyle{apj}

\end{document}